\documentclass[12pt]{article}
\usepackage{amsmath}
\usepackage{amssymb}
\usepackage{amsthm}
\usepackage{psfrag}
\usepackage{graphicx}
\usepackage{hyperref}

\newcommand{\be}{\begin{equation}}
\newcommand{\ee}{\end{equation}}
\newcommand{\beqa}{\begin{eqnarray}}
\newcommand{\eeqa}{\end{eqnarray}}

\newcommand\m{\mu}
\newcommand{\g}{\gamma}
\newcommand\D{\Delta}
\newcommand\n{\nu}
\renewcommand\l{\lambda}

\renewcommand\r{\rho}
\newcommand\s{\sigma}

\renewcommand\a{\alpha}
\renewcommand\b{\beta}
\newcommand\T{\Theta}
\newcommand{\HH}{{\cal H}}

\def\e{{\rm e}}
\def\d{\partial}
\newcommand{\bseq}{\begin{subequations}}
\newcommand{\eseq}{\end{subequations}}

\renewcommand{\ln}{\mathop{\rm ln}\nolimits}

\newcommand{\di}{\mathrm d}

\title{{\sc  Technically natural dark energy from Lorentz breaking}}

\author{D. Blas,\!$^a$ S. Sibiryakov,\!$^{b}$\vspace{.2cm}\\
\normalsize\llap{$^a$}
 \it FSB/ITP/LPPC,
 \'Ecole Polytechnique F\'ed\'erale de Lausanne,\\
 \normalsize\it CH-1015, Lausanne, Switzerland\\
\normalsize\llap{$^b$} \it Institute for Nuclear Research of the
Russian 
Academy of Sciences, \\
      \normalsize \it  60th October Anniversary Prospect, 7a, 
117312 Moscow, Russia}
\date{}

\begin{document}
\maketitle

\begin{abstract}
We construct a model of dark energy with a technically natural small contribution to cosmic acceleration, i.e. this
contribution does not receive corrections from other scales in the
theory. 
The proposed acceleration mechanism appears generically in the low-energy limit
of gravity theories 
with violation of Lorentz invariance that contain a derivatively
coupled scalar field $\Theta$. The latter 
may be the Goldstone field of a broken global symmetry. The model, that we call $\Theta$CDM, is a valid effective field theory up to a high cutoff just a few orders of magnitude below the Planck scale. Furthermore,  it can be 
ultraviolet-completed in the context of Ho\v rava gravity.
We discuss the observational predictions of the model.
Even in the  absence of a  cosmological constant term,
 the expansion history of the Universe is essentially
 indistinguishable from that of $\Lambda$CDM. 
 The difference between the two theories appears at the level of
 cosmological perturbations. 
 We find that in $\T$CDM the matter power spectrum is enhanced at
subhorizon scales compared to $\Lambda$CDM. 
This property can be used to discriminate the model from
$\Lambda$CDM with current cosmological data. 
\end{abstract}
\newpage
\tableofcontents

\section{Introduction}

Current cosmological observations 
indicate that the Universe recently entered a phase of accelerated 
expansion (see e.g. \cite{Durrer:2011gq} and references therein). 
 The driving force responsible for this phenomenon is normally called
 dark energy (DE) and its nature remains one of the major puzzles of 
contemporary physics. 
The standard explanation invokes a
cosmological constant (CC), and the corresponding model --
$\Lambda$CDM --
manages to 
fit the data rather well \cite{Durrer:2011gq}. However, on the
theoretical side this 
explanation faces a serious problem. 
The measured value of the CC is many orders of magnitude smaller than
the one expected on theoretical grounds.
The latter is estimated as the value of quantum corrections to
the vacuum energy 
plus the contribution related to possible phase transitions in the
early Universe  
and is set by the energy scales of elementary
particle interactions. This is the essence of the famous old Cosmological
Constant Problem
\cite{Weinberg:1988cp}. 

There exist two approaches to address this problem. The anthropic
explanation assumes that the CC is indeed
behind DE and claims that the observed value of the
CC is small because the very existence of observers is possible only
in a universe with small CC. The dynamical realization of this idea 
invokes the concept of a landscape of many vacua with
different values of the potential energy and transitions between
them. Thus the underlying theory must be necessarily very complex, and
it is a matter of ongoing debate whether  this picture possesses any
predictive power \cite{Starkman:2006at,Hartle:2007zv,Page:2007bt,
Garriga:2007wz,Bousso:2010vi}.
   
An alternative to the anthropic reasoning is to declare that DE is not related to the CC. 
For this mechanism to work, one first needs to get rid of the CC contribution to acceleration. 
One may speculate that there is some deep
(presumably nonperturbative)
mechanism that reduces the value of the CC, defined as the value of the
total potential energy of all the fields at the absolute minimum, to zero 
\cite{Linde:1988ws,Hawking:1984hk,Coleman:1988tj}. In this picture the
observed accelerated expansion of the Universe must either be produced 
by some dynamical component of the matter sector or be due to
modification of the laws of gravity at the cosmological distances
leading to theories of \emph{dynamical} DE 
(sometimes referred to as quintessence). Many models of this type
have been constructed, see \cite{Copeland:2006wr} for a
review. However, a lot of them suffer from one or a few of
common problems (which are often just other faces of 
the CC problem). The simplest models of quintessence involve very
light scalar fields with the mass smaller than the present Hubble
parameter. Generically, these masses are unstable with respect to
quantum corrections which reintroduces a fine-tuning problem. With the
additional assumption of an approximate global shift symmetry of the scalar
field (in other words, assuming that this field is a pseudo-Goldstone
boson) its mass can be protected from perturbative
corrections \cite{Frieman:1995pm}. However, the VEV that breaks the
symmetry corresponding to the pseudo-Goldstone field is required to
have the value of order or above the Planck scale,
which is believed to be problematic when
non-perturbative quantum gravity effects are taken into
account \cite{Kallosh:1995hi}. Finally, 
from the purely observational perspective a
drawback of
this last class of theories is that, except for fine-tuned cases, 
their predictions are practically
indistinguishable from those of $\Lambda$CDM which makes impossible
the discrimination of these theories from
the CC. From this perspective, the models
of DE involving non-linear potentials and/or kinetic terms for
the fields and their interaction with dark and ordinary matter 
are more interesting due to their non-trivial phenomenology.
However, the non-linear structure of the potentials makes these theories
strongly coupled at a low cutoff scale
(typically of order $10 $ eV). Thus, these models can be understood
only as effective theories below this cutoff, raising the issue of a
missing 
ultraviolet (UV)
completion at low energy scales.

In this paper we present a simple model for dynamical DE 
that avoids 
the aforementioned problems. Assuming a vanishing CC, the smallness of
the parameter setting the value 
of DE density in the model is protected by a discrete
symmetry. The model is a valid effective field theory up to a very
high scale just a few orders below 
the Planck mass. Remarkably, above this scale the model admits
a UV completion in the form of the Ho\v rava's model
\cite{Horava:2009uw} (more precisely, its consistent non-projectable
extension \cite{Blas:2009qj,Blas:2009ck,Blas:2010hb}) which is a
candidate for a renormalizable theory of quantum gravity. Finally,
while the expansion history of the Universe in our model is basically
indistinguishable from that of $\Lambda$CDM, the growth of
cosmological perturbations is different, which can allow to
discriminate this model from $\Lambda$CDM using the structure
formation and weak lensing
data\footnote{A non-trivial feature of the model is parametric
  enhancement of the signal in the structure formation that allows to
  have sizable effect even after satisfying stringent Solar system
  constraints.}. 

The price to pay for these nice features is to admit
violation of Lorentz symmetry in the gravity sector. This possibility
is interesting on its own right and has received a great deal of
attention recently, the prototypical models describing the possible
patterns of Lorentz symmetry breaking being the Einstein-aether 
\cite{Jacobson:2000xp} and
the ghost condensate \cite{ArkaniHamed:2003uy} theories.
A major issue in this kind of theories is the
possible transmission of Lorentz breaking to the fields of the
Standard Model (SM) which would be unacceptable as 
Lorentz invariance is tested to a very high precision in the SM
sector \cite{Kostelecky:2008ts}. 
Nevertheless, this problem 
is completely unrelated to
DE and may be addressed using some dynamical mechanisms
\cite{GrootNibbelink:2004za,Bolokhov:2005cj,Giudice:2010zb}. Once the 
emergence of Lorentz
invariance at low energies for the SM is achieved, 
there is no extra requirement to make our proposal for DE technically
natural.

It is worth mentioning 
another interesting property of the model we are going to present that
may shed some light on the old CC problem (vanishing of the vacuum
energy). 
In the absence of a CC and matter (including dark matter) it
admits {\it two} solutions, corresponding to Minkowski and de Sitter
space-times. The latter solution is stable while the former is
unstable with respect to long-wavelength perturbations. For that
reason, the Minkowski solution is never reached during the cosmological
expansion and instead the system approaches the de Sitter branch. In
this sense the cosmology of our model is self-accelerating
(cf. \cite{Gorbunov:2008dj}). The instability of the Minkowski vacuum
is a purely infrared phenomenon and is cut off at certain finite
momentum. In other words, the Minkowski vacuum is perfectly
well-defined from the point of view of the UV physics. This is
consistent with the previous idea that the  existence of such a vacuum may be required
by some unknown principle which assures the cancellation of
CC. In this paper we will not pursue the discussion about what
kind of principle this could be.

This paper is organized as follows. In Sec.~\ref{Sec:self} we present
the model and describe the cosmological solutions. In
Sec.~\ref{Sec:pertprev} we make a preliminary analysis of
perturbations in the expanding and Minkowski backgrounds and show that
in the latter case long-wavelength modes are unstable. In
Sec.~\ref{Sec:lineq} we turn to the detailed analysis of the cosmological
perturbations in the Friedmann -- Robertson -- Walker (FRW) universe
and derive the set of linearized equations. These equations are solved
analytically in Sec.~\ref{Sec:pertan} in various
approximations. The reader who is interested in the final results can
skip this section and go directly to 
Sec.~\ref{Sec:pertnum} where we report the results of the
numerical integration of the linearized
equations. Section~\ref{Sec:conclusions} is devoted to conclusions. 
Appendix \ref{App:A} contains some
details about our numerical procedure. 

\section{$\Theta$CDM model for cosmic acceleration}
\label{Sec:self}

We start by considering theories where local Lorentz invariance is broken by
the presence of a unit time-like vector $u_\m$. We will restrict to the
case where the vector is orthogonal to a set of space-like
3-dimensional surfaces that foliate the space-time. Parameterizing
these surfaces as the constant levels of a function $\varphi(x)$,
$u_\m$ can be written in the form\footnote{Our convention for the
  metric signature is  
$(+,-,-,-)$.}, 
\be
\label{udef}
u_\m=\frac{\d_\m\varphi}{\sqrt{g^{\nu\rho}\d_\n\varphi\d_\r\varphi}}\;.
\ee
Note that $u_\m$ is invariant under reparametrizations 
\be
\label{reparam}
\varphi\mapsto \tilde\varphi=f(\varphi),
\ee
for an arbitrary monotonic function $f$. The most general action
describing the coupling of $u_\m$ to gravity and containing a total of
two derivatives acting on $u_\m$ has the form\footnote{We use 
subindices inside square brackets to distinguish the quantities referring
to the different sectors of the model.},
\be
\label{khronoact}
S_{[EH\chi]}=-\frac{M_0^2}{2}\int \di^4x \sqrt{-g}
\big(R+K^{\m\n}_{\phantom{\m\n}\s\r}\nabla_\m u^\s\nabla_\nu u^\rho\big)\;,
\ee
where $R$ is the space-time curvature, $M_0$ is a mass parameter related to the Planck mass
and
\be
\label{Kmunulr}
K^{\mu\nu}_{\phantom{\m\n}\sigma\rho}=\beta\delta^\mu_\rho\delta^\nu_\sigma
+\lambda\delta^\mu_\sigma\delta^\nu_\rho
+\alpha u^\mu u^\nu g_{\sigma\rho}\;,
\ee 
with  $\a,\b,\l$ free dimensionless
parameters. This model has been proposed\footnote{Note that the
  parameter $\l$ in (\ref{khronoact}) corresponds to $\l'$ in the
  notations of \cite{Blas:2010hb}.} in 
\cite{Blas:2010hb} (see also \cite{Jacobson:2010mx}) and was dubbed
`khrono-metric' because it introduces the notion of a preferred global
time coordinate set by the field $\varphi(x)$ -- the `khronon' field. As discussed
in \cite{Blas:2010hb}, Eq.~(\ref{khronoact}) describes the low-energy
limit of the consistent extension \cite{Blas:2009qj,Blas:2009ck} of
Ho\v rava gravity \cite{Horava:2009uw}. The latter has been
proposed as a power-counting renormalizable model of quantum gravity.
In the case when the parameters $\a,\b,\l$ are of the same order 
the action (\ref{khronoact}) gives a valid effective field theory up
to the scale $\Lambda_{cutoff}=\min\{M_0\sqrt\a, M_0\}$. Note
  that when written in terms of the khronon field $\varphi$ the action
  (\ref{khronoact}) contains four derivatives and thus gives rise to a
  fourth order equation of motion. However, the structure of the
  higher derivative terms is such that,
 for a proper choice of the
  time variable, only two of these derivatives are temporal
  \cite{Blas:2010hb}. Thus the
  khronon field contains only one propagating degree of freedom. In
  particular, no instabilities, that are in general associated with
  higher derivative Lagrangians, arise.
   
Experimental data impose constraints on the parameters $\a,\b,\l$. For
general order one ratios between these constants the strongest bounds
come from the constraints on the 
parameters
$\a_1^{PPN},\a_2^{PPN}$ of the parametrized post-Newtonian (PPN)
formalism\footnote{These PPN parameters describe effects of local
  Lorentz violation and are zero in general relativity. The
  experimental bounds are \cite{Will:2005va}:
$ |\a_1^{PPN}|\lesssim 10^{-4},~|\a_2^{PPN}|\lesssim 10^{-7}\;.$}
and require $\a,\b,\l$ to be smaller than
$10^{-6}$ \cite{Blas:2010hb}. However, for the special case $\a=2\b$
both $\a_1^{PPN},\a_2^{PPN}$ vanish and one is left with a much weaker
bound $\a,\b,\l\lesssim 0.1$ coming from the dynamics of the Big Bang
Nucleosynthesis (BBN) \cite{Carroll:2004ai} and the emission of
gravitational waves in binary systems \cite{BlasSanc}. 
Another special case is $\b=0$, $\a=\l$. Then the PPN parameter
$\a_2^{PPN}$ vanishes, while the constraint on $\a_1^{PPN}$ gives
$\a,\l<10^{-4}$.   

If we do not insist on the existence of a known UV-completion and are
content with an effective theory with high cutoff,
we can relax the condition (\ref{udef}) and consider a general unit
time-like vector. This would lead us to the Einstein-aether theory
\cite{Jacobson:2000xp} (see \cite{Jacobson:2008aj} for a review). The
action for the general aether has the same form as (\ref{khronoact})
with an extra term\footnote{For a hypersurface-orthogonal
  aether this term can be written as a combination of 
those  present in (\ref{khronoact}).}.
Physically, the Einstein-aether model differs from the khrono-metric
theory by the presence of transverse vector modes among the
perturbations. However, these modes do not affect the homogeneous
cosmological expansion, nor the linear evolution of scalar
cosmological perturbations\footnote{The transverse modes can lead to
  interesting effects in the CMB polarization \cite{Nakashima:2011fu}.}. 
Thus, the analysis of the present paper applies
without changes to the general aether model. Just for clarity we
concentrate on the hypersurface-orthogonal case corresponding to the
khrono-metric theory.

To the system (\ref{khronoact}) we add a scalar field $\Theta$ with
the symmetry under the shifts
\be
\label{shiftsym}
\Theta\mapsto \Theta+const\;.
\ee
The precise origin of this field is not important for our
purposes. However, an
example that is useful to keep in mind is a Goldstone boson corresponding
to a spontaneously broken global symmetry\footnote{Note that the scale
of this putative symmetry breaking will be completely irrelevant for
the analysis in this paper (provided that it is higher than the
temperature of the universe at the epochs of interest, which is
roughly a few eV). In particular, it can be safely below the Planck mass.}.
For this reason, we will refer to the field $\Theta$ as
``Goldstone field'' in the rest of the paper. In contrast to the
pseudo-Goldstone 
quintessence models \cite{Frieman:1995pm}, we assume the symmetry
(\ref{shiftsym}) to be 
exact. The interaction between the Goldstone and the khronon must be
invariant under (\ref{reparam}) and (\ref{shiftsym}). Keeping only the
operators that have dimensions up to 4 and thus are relevant for 
physics at long distances, we obtain the following general
action\footnote{ One could consider including 
in the action operators 
\[
u^\m\nabla_\mu u^\n\d_\n \T~,~~~~ (\nabla_\m u^\m) u^\n\d_\n\T\;,
\]
that formally have dimension 3. We omit them for two reasons. 
First, these operators can be
forbidden by imposing the symmetry under simultaneous change of the signs of
$\T$ and $u^\m$: $\T\mapsto -\T\;,~u^\m\mapsto -u^\m$. More
importantly, these operators actually have dimension higher than 4 when
expressed in terms of the canonically normalized khronon
perturbations, and thus do not affect the long distance physics. 
} for
$\Theta$: 
\be
\label{Thetaact}
S_{[\Theta]}=\int \di^4x \sqrt{-g} \bigg(\frac{g^{\m\n}\d_\m\Theta\d_\n\Theta}{2}+
\varkappa\frac{(u^\mu\d_\mu\Theta)^2}{2}
+\mu^2u^\mu\d_\mu\Theta\bigg)\;.
\ee
Here the coupling constant $\varkappa$ is dimensionless, while $\mu$
has dimension of mass. It is important to notice that the $\m$-term is
the only one that is not 
invariant under $\Theta\mapsto -\Theta$. Thus, assuming that the UV completion of the theory
respects this discrete symmetry, the value of
$\mu$ is stable under radiative corrections.

The physical effect of
the two khronon--Goldstone couplings in (\ref{Thetaact}) is quite
different. Let us consider the dynamics of the Goldstone field in the background with
flat metric and $u_\m=(1,0,0,0)$. Then the third term in
(\ref{Thetaact}) becomes a total derivative and does not affect the
dynamics, while the second term modifies
the propagation velocity $c_\Theta$ of the Goldstone field:
\[
c_\Theta^2=\frac{1}{1+\varkappa}\;.
\]
We will see in a moment that the effects of the khronon--Goldstone
coupling described by the third term become essential in curved
backgrounds and modify the dynamics of both fields at
cosmological scales. A coupling of this form was introduced in
\cite{Donnelly:2010cr} in the context of inflation in the
Einstein-aether model. A similar coupling was also considered in
\cite{Libanov:2007mq}. 

We will study the cosmological evolution in the model with the action
\be
\label{Acts}
S=S_{[EH\chi]}+S_{[\Theta]}+S_{[m]}\;,
\ee
where $S_{[EH\chi]}$ and $S_{[\Theta]}$ have been introduced above and
$S_{[m]}$ stands for the 
action of matter, which includes baryonic matter, dark matter and radiation. We assume that $S_{[m]}$ has the
standard form with matter interacting minimally with the metric
$g_{\m\n}$. In particular, we assume that there are no direct
couplings between the matter sector and the khronon or Goldstone
fields\footnote{This must be considered as a simplifying
  assumption. For the ordinary matter the direct couplings to the
  khronon field would imply violation of the Lorentz
  symmetry within the SM sector and is tightly constrained
  by the experimental data. Once these constraints are satisfied, the
  effects in cosmology are negligible.
  On the other hand, for dark matter 
we do not have such strong
  observational evidence of Lorentz invariance. 
Allowing for direct coupling between dark matter and the khronon
  may lead to interesting effects that will be reported elsewhere 
\cite{BSinprep}. As for the coupling of SM / dark matter to the
Goldstone, due to the symmetry (\ref{shiftsym}) it must involve
derivatives. Apart from possible kinetic mixing with a scalar gauge
singlet that may be part of dark matter, this would produce higher
order operators irrelevant for long-distance physics.}.     
We will refer to this model as $\T$CDM. 

Introducing the 
Ansatz for the most general spatially homogeneous solution
\bseq
\begin{gather}
\label{cosmetr}
\di s^2=N^2(t)\di t^2-a^2(t) \delta_{ij}\di x^i \di  x^j\;,\\
\label{cosmfields}
\varphi=\varphi(t)\;,~~~\Theta=\Theta(t),
\end{gather}
\eseq
into the actions (\ref{khronoact}), (\ref{Thetaact}) we obtain
\[
S_{[EH\chi]}+S_{[\Theta]}=-\frac{M_0^2(6+3\beta+9\lambda)}{2}\int \di^4x\frac{\dot a^2a}{N}
+\int
\di^4x\bigg(\frac{a^3}{2Nc_\Theta^2}\dot\Theta^2+\mu^2a^3\dot\Theta\bigg)\;. 
\]
Note that the field $\varphi$ has dropped out of the action, which is a consequence of the 
invariance under (\ref{reparam}). Varying
the action with respect to $\Theta$ and $N$ we obtain the cosmological
equations,
\begin{align}
\label{Phicosmeq}
&\frac{\di}{\di t}\bigg(\frac{a^3}{Nc_\T^2}\dot\T+\mu^2 a^3\bigg)=0\;,\\
\label{Fried}
&H^2=\frac{8\pi G_{cosm}}{3}\bigg(\frac{\dot\T^2}{2N^2c_\T^2}+\rho_{m}\bigg)\;,
\end{align}
where $H\equiv{\dot a}/{(a N)}$ is the Hubble rate and
\[
G_{cosm}=\frac{1}{8\pi M_0^2(1+\beta/2+3\lambda/2)}\;.
\]
On the r.h.s. of the
Friedmann equation (\ref{Fried}) we added 
the contribution $\rho_{m}$ of
ordinary matter. 
Note that the expression for the gravitational
constant $G_{cosm}$ entering the Friedmann equation is corrected by
the presence of the khronon compared to the 
standard GR formula
$G=(8\pi M_0^2)^{-1}$. As discussed in \cite{Blas:2009qj,Blas:2010hb},
this differs from the expression for the
gravitational coupling 
\[
G_{N}=\frac{1}{8\pi M_0^2(1-\a/2)},
\]
appearing in the Newtonian potential of a localized source.
The bound \cite{Carroll:2004ai} 
$|G_{cosm}/G_N-1|\leq 0.13$ from BBN
thus requires the parameters $\a,\b,\l$ to be smaller
than $0.1$. 

The equation (\ref{Phicosmeq}) is easily solved with the result 
\be
\label{Phicosm}
\frac{\dot\T}{N}=-\bigg(\mu^2c_\T^2+\frac{C}{a^3}\bigg)\;,
\ee
where $C$ is an integration constant. Due to the cosmological expansion 
the ratio $\dot\T/N$ is
attracted to the constant non-zero value $-\mu^2c_\T^2$. On this
attractor solution the field $\T$ produces the same contribution into the
Friedmann equation (\ref{Fried}) as
the vacuum energy density 
\be
\label{vaceff}
V_{0,\,eff}=\mu^4c_\T^2/2\;.
\ee
In this way the model leads
to accelerated cosmology even in the absence of the true cosmological
constant. 
Due to the stability of  
the parameter $\mu$ under radiative corrections it is technically
natural for the effective vacuum energy density (\ref{vaceff}) to be small.  

Notice that the general solution (\ref{Phicosm}), once introduced in 
 the Friedmann equation (\ref{Fried}), also produces a contribution
 scaling like $a^{-3}$, that one may be tempted to interpret as dark
 matter. However, this contribution is always accompanied by the stiff
 term proportional to $C^2a^{-6}$ that dominates over the dark 
matter-type contribution. To comply with the experimental constraints the
 stiff term must be small. We will set $C=0$ in what follows.
Thus the $\T$-field cannot be responsible for dark matter, which one
has to add as a separate component. This
 will be included in the part of the theory
 described by $S_{[m]}$ in~(\ref{Acts}). 

It is worth stressing that we do not present any mechanism that could
lead to vanishing of the vacuum energy and thus do not pretend to
solve the old CC problem. However, we point out the following
interesting property of the $\T$CDM model that may shed some light on
the resolution of
this problem. 
In the absence of matter, $\rho_{m}=0$, the system of
equations (\ref{Phicosmeq}), (\ref{Fried}) possesses a static solution
with flat metric,
\be
\label{Mink}
N=a=const,\;~~~\dot\T=0\;.
\ee
However, this solution is
never achieved if the evolution starts from a 
configuration with $\rho_{m}\neq 0$. Moreover, in the next section
we will see that the Minkowski solution is unstable with respect to
long-wavelength perturbations. On the other hand, at small scales the
instability is cut off and from the viewpoint
of the short distance physics (\ref{Mink}) is a valid background. 
Thus one may hope to find a mechanism that would neutralize the CC in
this background and ensure the existence of a (unstable) Minkowski
vacuum. 
However, the dynamics drives the universe not to Minkowski
but to the de Sitter attractor.

\section{Preview of linear perturbations}
\label{Sec:pertprev}

As we saw in the previous section, at the level of homogeneous
cosmology the attractor solution in the  $\T$CDM is
indistinguishable from $\Lambda$CDM. Then it is natural to wonder if
the model leaves any distinctive signatures at the next level of
approximation, namely in the evolution of linear cosmological
perturbations. The analysis of the linearized perturbations is also
indispensable to verify the stability of the accelerated
solution. 

Before going to the complete analysis which we postpone to the
subsequent sections, let us consider a simplified setup where one
neglects back-reaction of the khronon and Goldstone perturbations on the
geometry. Formally this corresponds to taking the limit
$M_0\to\infty$. As will be clear later, we must at the same time keep
the scale
\be
\label{Malpha}
M_\a\equiv\sqrt\a M_0
\ee 
fixed. This implies that in the limit of interest $\a\to 0$. We also
assume that the ratios $\b/\a$ and $\l/\a$ are fixed and of order one
in this limit.

We start with the analysis of perturbations in Minkowski
space-time. One writes
\bseq
\label{sigTetpert}
\begin{align}
\label{sigpert}
&\varphi(t,{\bf x})=\bar\varphi(t)+\chi(t,{\bf x})\;,\\
\label{Tetpert}
&\T(t,{\bf x})=\bar\T(t)+\xi(t,{\bf x})\;,
\end{align} 
\eseq
where $\bar\varphi(t)$, $\bar\T(t)$ are the background values of the
fields and $\chi$, $\xi$ are the perturbations. Note that due to the
reparametrization symmetry (\ref{reparam}) we can always set 
\be
\label{sigmagauge}
\bar{\varphi}(t)=t\;.
\ee
Also for the Minkowski solution $\bar\T(t)=const$. Expanding the
actions (\ref{khronoact}), (\ref{Thetaact}) to quadratic order in
fluctuations we obtain
\be
\label{S2}
[S_{[EH\chi]}+S_{[\Theta]}]^{(2)}=\int \di^4x\bigg[\frac{M_\alpha^2}{2}(\d_i\dot\chi)^2
-\frac{M_\a^2c_\chi^2}{2}(\D\chi)^2
+\frac{\dot\xi^2}{2c_\T^2}-\frac{(\d_i\xi)^2}{2}
-\m^2\d_i\chi\d_i\xi\bigg]\;,
\ee
where $\D\equiv\d_i\d_i$ is the spatial Laplacian, and we have
introduced the notation
\be
\label{sound}
c_\chi^2=\frac{\b+\l}{\alpha}\;.
\ee 
Throughout the paper we will assume that $c_\chi^2$, as well as
$c_\T^2$, are of order one.
The action (\ref{S2}) yields the equations of motion,
\bseq
\label{coupledMink}
\begin{align}
\label{coupledchiMink}
&\D\bigg(\ddot\chi-c_\chi^2\Delta\chi
+\frac{\mu^2}{M_\alpha^2}\xi\bigg)=0\;,\\
\label{coupledxiMink}
&\frac{\ddot\xi}{c_\T^2}-\Delta\xi-\mu^2\Delta\chi=0\;.
\end{align}
\eseq
The overall Laplacian in the first equation can be cancelled out for
any inhomogeneous configuration. Performing the Fourier decomposition
\be
\label{Fourier}
\chi,~\xi\propto\e^{-i\omega t+i{\bf kx}}
\ee
we find the dispersion relations of the propagating modes,
\be
\label{dispMink}
\omega^2_\pm=\frac{1}{2}\bigg[(c_\chi^2+c_\T^2)k^2\pm
\sqrt{(c_\chi^2-c_\T^2)^2 k^4
+\frac{4\mu^4c_\T^2}{M_\alpha^2}k^2}\bigg]\;.
\ee
This expression simplifies in two regimes. At $k$ larger than the
critical momentum
\be
\label{kc}
k_c\equiv \m^2/M_\a\;,
\ee
the fields $\chi$ and $\xi$ decouple and describe two modes with linear
dispersion relations
\bseq
\label{displarge}
\begin{align}
\label{ochilarge}
&\omega^2_\chi=c_\chi^2k^2\;,\\
\label{oxilarge}
&\omega^2_\xi=c_\T^2k^2\;.
\end{align}
\eseq
Note that both modes are stable.
In the opposite case $k\ll k_c$ Eq.~(\ref{dispMink}) takes the form, 
\be
\label{Minkowsk}
\omega^2_\pm=\pm c_\T k_c k\;.
\ee 
This dispersion relation is non-analytic: the frequency
is proportional to the square root of the momentum. Clearly, one of
the modes has purely imaginary frequency and thus exponentially grows
with time. This signals instability of the Minkowski background with
respect to long-wavelength perturbations. As we are going to see, 
this instability disappears in the expanding universe.

As the next exercise we consider the evolution of the $\chi$-,
$\xi$-perturbations in an external FRW metric. We will work in the
conformal time which corresponds to setting $N=a$ in
(\ref{cosmetr}), and consider a general time-dependence of the scale
factor. We expand again as in (\ref{sigTetpert}) where we fix
(\ref{sigmagauge}). The background configuration $\bar\T(t)$ satisfies
Eq.~(\ref{Phicosm}). To simplify the calculations we concentrate on
the attractor solution 
\be
\label{attr}
\dot{\bar\T}=-a\m^2c^2_\T\;.
\ee 
After a straightforward computation we find the quadratic action
\be
\label{S2FRW}
\begin{split}
[S_{[EH\chi]}+S_{[\Theta]}]^{(2)}=\int \di^4x\,a^2\,\bigg[&\frac{M_\alpha^2}{2}(\d_i\dot\chi)^2
-\frac{M_\a^2c_\chi^2}{2}(\D\chi)^2\\
&-\big(M_\a^2\dot\HH(1-B)+M_\a^2\HH^2(1+B)+a^2\m^4c_\T^4\big)
\frac{(\d_i\chi)^2}{2}\\
&+\frac{\dot\xi^2}{2c_\T^2}-\frac{(\d_i\xi)^2}{2}
-a\m^2 c_\T^2\d_i\chi\d_i\xi\bigg]\;,
\end{split}
\ee
where we introduced the notations
\begin{align}
\label{Bdef}
&B=\frac{\b+3\l}{\a}\;,\\
\label{HHdef}
&\HH\equiv\frac{\dot a}{a}=Ha\;,
\end{align}
Note the appearance of a term with two space derivatives of $\chi$
which was absent in the case of Minkowski background. The action \eqref{S2FRW}
yields the following equations,
\bseq
\label{coupled}
\begin{align}
\label{coupledchi}
&\ddot\chi+2\HH\dot\chi-c_\chi^2\Delta\chi
+\big[\dot\HH (1-B)+\HH^2(1+B)+a^2c_\T^4k_c^2\big] \chi
+\frac{a c_\T^2k_c^2}{\m^2}\xi=0\;,\\
\label{coupledxi}
&\ddot\xi+2\HH\dot\xi-c_\T^2\Delta\xi-a\mu^2c_\T^4\Delta\chi=0\;,
\end{align}
\eseq
where in the first equation we have cancelled the overall
Laplacian. The crucial difference from Eqs.~(\ref{coupledMink}) 
in the Minkowski case is the presence of an
effective mass term for the field~$\chi$, 
\be
\label{meff}
m_{eff}^2=\frac{\dot\HH(1-B)+\HH^2(1+B)}{a^2}
+c_\T^4 k_c^2\;.
\ee 
This expression deserves two comments. First, during the epoch of primordial
inflation one can neglect the last term in (\ref{meff}) and also use
$\dot\HH\approx\HH^2$. 
This gives $m_{eff}^2\approx 2H_{inf}^2$, i.e. the effective khronon mass
is of order the Hubble parameter at inflation and coincides with the
effective mass 
of a conformally coupled
scalar  field. 
As a consequence, the khronon perturbations are not
generated during inflation.
This agrees with the result of \cite{Kobayashi:2010eh} that
no isocurvature perturbations associated with the khronon are produced
during inflation. 

Second, for certain choices of parameters the
effective mass square (\ref{meff}) can become negative during the
radiation and matter dominated epochs. This will lead to the growth of
the khronon perturbations with wavelengths larger than the horizon
size at the corresponding epoch. It is worth stressing that this
growth is under control, its rate being proportional to the Hubble
parameter. Even if this regime is definitely interesting from the
phenomenological point of view, it corresponds to a rather special
corner of the parameter space. In the present work 
we focus on the generic stable case.

Let us now analyze the behavior of subhorizon
modes with large momenta and frequencies, $k,\,\omega\gg\HH$. 
In this case we can neglect all
$\HH$-dependent terms in Eqs.~(\ref{coupled}) and perform the Fourier
decomposition (\ref{Fourier}). However, the presence of a non-trivial background
$\dot{\bar\T}$ still modifies the equations for the perturbations,
eliminating any instability  for the momenta inside the horizon. 
Explicitly, the dispersion
relations are:\footnote{Here $\omega$ and $k$ should be understood as
  physical frequency and momentum which corresponds to putting $a=1$
  in the formulas.}
\[
\omega^2_\pm=\frac{1}{2}\Big[(c_\chi^2+c_\T^2)k^2
+c_\T^4 k_c^2
\pm \sqrt{\big((c_\chi^2+c_\T^2)k^2
+c_\T^4 k_c^2\big)^2
-4c_\chi^2 c_\T^2 k^4}\Big]\;.
\]
For large momenta $k\gg k_c$ one
recovers the linear dispersion relations (\ref{displarge}). While at
$k\ll k_c$ we have 
\begin{align}
\label{omega1}
&\omega_+^2=c_\T^4k_c^2+(c_\chi^2+c_\T^2)k^2\;,\\
\label{omega2}
&\omega_-^2=\frac{c_\chi^2k^4}{c_\T^2k_c^2}\;.
\end{align}
We see that, in contrast to the case of Minkowski, both modes are
stable (cf. (\ref{Minkowsk})). One of them possesses a frequency gap, while the other has a
dispersion relation with quadratic dependence of the frequency on
momentum. This implies that the latter mode has low propagation
velocity $c_{-}\sim k/k_c\ll 1$ and one expects it 
to cluster in the gravitational potential wells 
enhancing the growth of the cosmological perturbations. We are going to
see below that this expectation is correct. 

It is worth stressing that the analysis of this, as well as the subsequent
  sections, applies without
  change to the longitudinal sector of linear perturbations in the
  case when the khrono-metric model is substituted by the
  Einstein-aether theory to describe the Lorentz breaking.

\section{Linearized equations in FRW}
\label{Sec:lineq}

We now turn to the derivation of linearized equations of motion for
$\T$CDM including gravitational perturbations and matter fields. We start with the
Einstein's equations
\be
\label{Eins}
R_{\m\n}-\frac{1}{2}g_{\m\n}R=\frac{1}{M_0^2}
(T_{[\chi]\m\n}+T_{[\T]\m\n}+T_{[m]\m\n})\;,
\ee 
where $T_{[\chi]\m\n}$, $T_{[\T]\m\n}$ and $T_{[m]\m\n}$ are respectively the
energy-momentum tensors of the khronon, Goldstone field and matter. They
are defined as the variation of the corresponding action with respect
to the metric,
\[
T_{[i]\mu\nu}=\frac{2}{\sqrt{-g}}\frac{\delta S_{[i]}}{\delta g^{\mu\nu}}\;.
\]
After a long but straightforward computation we obtain
\be
\label{Tkh}
\begin{split}
M_0^{-2}T_{[\chi]\mu\nu}=-\nabla_\lambda K^{\lambda}_{\phantom{\lambda}\rho} u^\rho
u_{\mu}u_{\nu} +2\nabla_\lambda K^{\lambda}_{\phantom{\lambda}(\mu} u_{\nu)}
-\nabla_\lambda\big(K^\lambda_{\phantom{\lambda}(\mu}u_{\nu)}\big)
-\nabla_\lambda\big(K_{(\mu\nu)}u^\lambda\big)
\\
+\nabla_\lambda\big(K_{(\mu}^{\phantom{\m}\lambda} u_{\nu)}\big) +\alpha\big[a_\lambda a^\lambda u_\mu u_\nu
-2a^\lambda\nabla_{(\mu} u_\lambda u_{\nu)}
+a_\mu a_\nu\big]
+\frac{1}{2}K^\lambda_{\phantom{\lambda}\rho}\nabla_\lambda u^\rho g_{\mu\nu}\;,
\end{split}
\ee 
where
\bseq
\label{aKdef}
\begin{gather}
\label{amu}
a_\mu=u^\lambda\nabla_\lambda u_\mu\;,\\
\label{Kmunu}
K^\mu_{\phantom{\m}\nu}=K^{\lambda\mu}_{\phantom{\lambda\r}\rho\nu}\nabla_\lambda u^\rho\;,
\end{gather}
\eseq
and the round brackets denote symmetrization of indices:
\[
K_{(\mu\nu)}=\frac{1}{2}(K_{\mu\nu}+K_{\nu\mu})\;.
\]
Similarly, for $T_{[\Theta]\m\n}$ we find,
\be
\label{TPhi}
\begin{split}
T_{[\T]\mu\nu}=\d_\mu\T\d_\nu\T
+2\big(\varkappa\,u^\lambda\d_\lambda\T+\mu^2\big)\,u_{(\mu} \d_{\nu)}\T
-\,u_\mu u_\nu\big(\varkappa(u^\lambda\d_\lambda\T)^2+
\mu^2 u^\lambda \d_\lambda\T\big)\\ 
-\bigg(\frac{\nabla^\lambda\T\d_\lambda\T}{2}
+\frac{\varkappa}{2}(u^\lambda\d_\lambda\T)^2
+\mu^2u^\lambda\d_\lambda\T\bigg)g_{\mu\nu}\;.
\end{split}
\ee

Finally, for the matter sector we 
will include only two decoupled components corresponding to cold
matter (cm) and
radiation ($\gamma$) in the perfect fluid approximation. Reducing the matter
sector to this form is, of course, a crude simplification but it is 
convenient for illustrating the difference
between $\T$CDM and $\Lambda$CDM. In this respect, whenever we
refer to the $\Lambda$CDM model in the future, we will have in mind the 
case of GR with a CC and the previous matter content.  We leave 
more accurate analysis of the cosmological perturbations for 
future research. Thus, for the energy-momentum tensor we take the standard
hydrodynamic expression for each component
\be
\label{Tmat}
T_{[m]\m\n}=(\r_{[\g]}+p_{[\g]})v_{[\g]\m} v_{[\g]\n}-p_{[\gamma]} g_{\m\n}+(\r_{[cm]}+p_{[cm]})v_{[cm]\m}v_{[cm]\n}\;,
\ee
where $\r_{[a]}$, $p_{[a]}$ are the energy density and pressure of the
different matter components and 
 $v_{[a]\m}$ are their 4-velocities.

To form a closed system the Einstein's equations must be supplemented by the
equations of motion for the khronon, Goldstone field and matter. The
first two of these equations are obtained by varying the sum of the actions
(\ref{khronoact}) and (\ref{Thetaact}) with respect to $\varphi$ and
$\T$. This yields,
\bseq
\label{khTeteq}
\begin{align}
\label{kheq}
\nabla_\rho\bigg[\frac{P^{\rho\mu}}{\sqrt{X}}
\bigg(-\nabla_\nu K^\nu_{\phantom{\n}\mu}
+\alpha\,a_\nu\nabla_\mu u^\nu 
-\frac{1}{M_0^2}
\d_\mu\T(\varkappa\,u^\lambda\d_\lambda\T+\mu^2)\bigg)\bigg]=0\;,\\
\label{Teteq}
\Box\T+\varkappa\nabla_\mu(u^\mu u^\nu\d_\nu\T)+\mu^2\nabla_\mu u^\mu=0\;,
\end{align}
\eseq
where 
\begin{gather}
P^{\mu\nu}=g^{\mu\nu}-u^\mu u^\nu\;,\notag\\
X=g^{\m\n}\d_\mu\varphi\d_\nu\varphi\;,\notag\\
\Box=g^{\m\n}\d_\m\d_\n\;.\notag
\end{gather}
The equations of motion for matter follow from the conservation
of the energy-momentum tensor of each component (recall that we assume
 the two components to be decoupled),
\be
\label{eqmat}
\nabla^\m T_{[\g]\m\n}=\nabla^\m T_{[cm]\m\n}=0\;.
\ee
In fact, only one of the equations in (\ref{eqmat}) is an independent
equation, as the conservation 
of the total matter energy-momentum tensor follows from the Einstein's
equations and 
(\ref{khTeteq}).

We now expand the above expressions to linear order in perturbations
around a FRW background. The dynamics differ from the standard case
only in the scalar sector of the perturbations and we concentrate on
this in what follows.
For the khronon and Goldstone fields we take the
representation (\ref{sigTetpert}) with the background values
satisfying (\ref{sigmagauge}), (\ref{attr}). The metric
perturbations are chosen in the conformal Newton's gauge
\[
\di s^2=a^2(t)\big[(1+2\phi)\di t^2-(1-2\psi)\delta_{ij}\di x^i \di x^j\big]\;.
\]
Finally, for the scalar sector of the matter perturbations we write
\[
\r_{[a]}=\bar\r_{[a]}(t)+\delta\r_{[a]}\;,~~~~
p_{[a]}=\bar p_{[a]}(t)+\delta p_{[a]}\;,~~~~
v_{[a]j}=\d_j v_{[a]}\;.
\]
Substituting these expressions into Eqs.~(\ref{Eins}), (\ref{Tkh}),
(\ref{TPhi}), (\ref{Tmat}) we obtain the expressions for the linearized Einstein's equations in $\T$CDM,
\bseq
\label{Einslin}
\begin{align}
2\D\psi-3\HH(2+\a B)\dot\psi-\a\D\phi
-\frac{2(\bar\r_{[\g]}+\bar\r_{[cm]})a^2}{M_0^2}\phi
+\a\D\dot\chi+\a(1-B)\HH\D\chi&
\notag\\
+\frac{\m^2a}{M_0^2}\dot\xi
-\frac{a^2(\delta\r_{[\g]}+\delta\r_{[cm]})}{M_0^2}&=0\;,
\label{Eins00}\\
\label{Eins0i}
(2+\a B)(\dot\psi+\HH\phi)+(\b+\l)\D\chi-\frac{(\bar\r_{[\g]}+\bar
  p_{[\g]})a}{M_0^2}v_{[\g]}-\frac{\bar\r_{[cm]}a}{M_0^2}v_{[cm]}&=0\;,\\
3(2+\a B)\big[\ddot\psi+\HH (\dot\phi+2\dot\psi)+(2\dot\HH+\HH^2)\phi\big]
+2\D\phi-2\D\psi\notag\\
+\a B(\D\dot\chi+2\HH\D\chi)-\frac{3a^2\delta p_{[\g]}}{M_0^2}&=0\;,
\label{Einsij1}\\
\label{Einsij2}
\phi-\psi-\beta(\dot\chi+2\HH\chi)&=0\;,
\end{align}
\eseq
where $B$ and $\HH$ are defined in (\ref{Bdef}), (\ref{HHdef}). 
These equations are respectively 
the $(00)$, $(0i)$, the trace and trace-free parts of the $(ij)$ 
Einstein's equations. In deriving (\ref{Einsij1}) we have used that the 
background metric satisfies the equation 
\[
(2+\a B)(2\dot\HH+\HH^2)-\frac{\m^4c_\T^2a^2}{M_0^2}
+\frac{2\bar p_{[\g]} a^2}{M_0^2}=0\;,
\]
which can be obtained by taking the time derivative of \eqref{Fried}.
Note that according to Eq.~(\ref{Einsij2}) for non-zero $\beta$ 
the khronon field induces anisotropic stress and thus the
gravitational potentials $\phi$ and $\psi$ are in general different.  

The equations of motion (\ref{khTeteq}) for the khronon and Goldstone field
read at the linear order:
\bseq
\label{khTetlin}
\begin{align}
\label{khlin}
\ddot\chi+2\HH\dot\chi-c_\chi^2\D\chi
+\Big(\dot\HH (1-B)+\HH^2(1+B)+&a^2c_\T^4k_c^2\Big)\chi
+\frac{a c_\T^2k_c^2}{\m^2}\xi
\notag\\
&=\dot\phi+\HH(1+B)\phi+B\dot\psi\;,\\
\label{xilin}
\ddot\xi+2\HH\dot\xi-c_\T^2\D\xi-a\m^2c_\T^4\D\chi
=-a\m^2c_\T^2\dot\phi-3a&\mu^2c_\T^2\HH\phi\;,
\end{align}
\eseq
where we remind the definitions (\ref{sound}), (\ref{kc}) of
$c_\chi$ and $k_c$. Note that these equations differ from
Eqs.~(\ref{coupled}) valid in an external FRW universe by the appearance of the
source term proportional to the metric perturbations on the r.h.s.

The perturbed matter equations coming from (\ref{eqmat}) 
have the standard
form:
\bseq
\label{matter} 
\begin{align}
\dot{\delta\r_{[a]}}+3\HH(\delta\r_{[a]}+\delta p_{[a]})-(\bar\r_{[a]}+\bar p_{[a]})
\left(\frac{\D v_{[a]}}{a}+3\dot\psi\right)=0\;,\\
\dot v_{[a]}+3\HH v_{[a]}+\frac{\dot{\bar\r}_{[a]}+\dot{\bar p}_{[a]}}{\bar\r_{[a]}+\bar p_{[a]}}v_{[a]}
-\frac{a\delta p_{[a]}}{\bar\r_{[a]}+\bar p_{[a]}}-a\phi=0\;.
\end{align}
\eseq
In the next sections we analyze the system of linear equations
(\ref{Einslin}), (\ref{khTetlin}), (\ref{matter}). 

\section{Cosmological perturbations: qualitative analysis}
\label{Sec:pertan}

We begin with a qualitative discussion of the evolution of
cosmological perturbations in the $\T$CDM model at radiation and matter
domination. From Eqs.~(\ref{Einslin}) it is easy to see that at these
stages 
the modifications of the linearized
equations of $\T$CDM, 
as compared to  the  $\Lambda$CDM case, are proportional to the
dimensionless parameters $\a,\b,\l$. As discussed in
Sec.~\ref{Sec:self}, consistency with the experimental data requires
these parameters to be small, $\a,\b,\l\ll 1$. This allows to solve
the system (\ref{Einslin}), (\ref{khTetlin}), (\ref{matter}) with the
following 
perturbative scheme. One first solves Eqs.~(\ref{Einslin}),
(\ref{matter}) in the 
zeroth-order approximation neglecting all the contributions
proportional to $\a,\b,\l$. Assuming  adiabatic
initial conditions, the solution is given by the standard
adiabatic mode. For clarity we concentrate on the modes entering
inside the horizon at the matter-dominated epoch. In this case the
adiabatic mode is particularly simple: the two gravitational
potentials $\phi$ and $\psi$ are equal and constant during radiation-
and matter-dominated epochs, with a jump at the radiation--matter
equality,
\be
\label{phiRDMD}
\psi^{(0)}=\phi^{(0)}=
\begin{cases}
\phi_\gamma=const\,,&\text{radiation domination};\\
\phi_{cm}=\frac{9}{10}\phi_\gamma\,,&\text{matter domination}.
\end{cases}
\ee
At the second step of the perturbative scheme these expressions are
substituted into 
Eqs.~(\ref{khTetlin}) as the source for the fields $\chi$ and $\xi$.  
Finally, the solution for $\chi$ and $\xi$ is inserted back into
Eqs.~(\ref{Einslin}) to produce the corrections $\phi^{(1)}$,
$\psi^{(1)}$ and $(\delta\rho_{[a]}/\r_{[a]})^{(1)}$ 
to the observable quantities. Thus our task at the moment is to solve
Eqs.~(\ref{khTetlin}) with the sources (\ref{phiRDMD}).

 A complete analysis should also
include the evolution at the epoch of DE-domination. However, as this
epoch started only recently it does not qualitatively affect the
results. At the same time, all the following analytic estimates should be
understood as giving only the qualitative picture. For example, they do not
capture the transient behavior at the boundaries of the different
dynamical regimes, which may be important due to finite duration of
the corresponding epochs. We will neglect these effects in the
rest of this section, and postpone their proper treatment to the numerical analysis (see Sec.~\ref{Sec:pertnum}).

\subsection{Different regimes of evolution}

\begin{figure}[!htb]
\begin{center}
\includegraphics[width=0.6\textwidth]{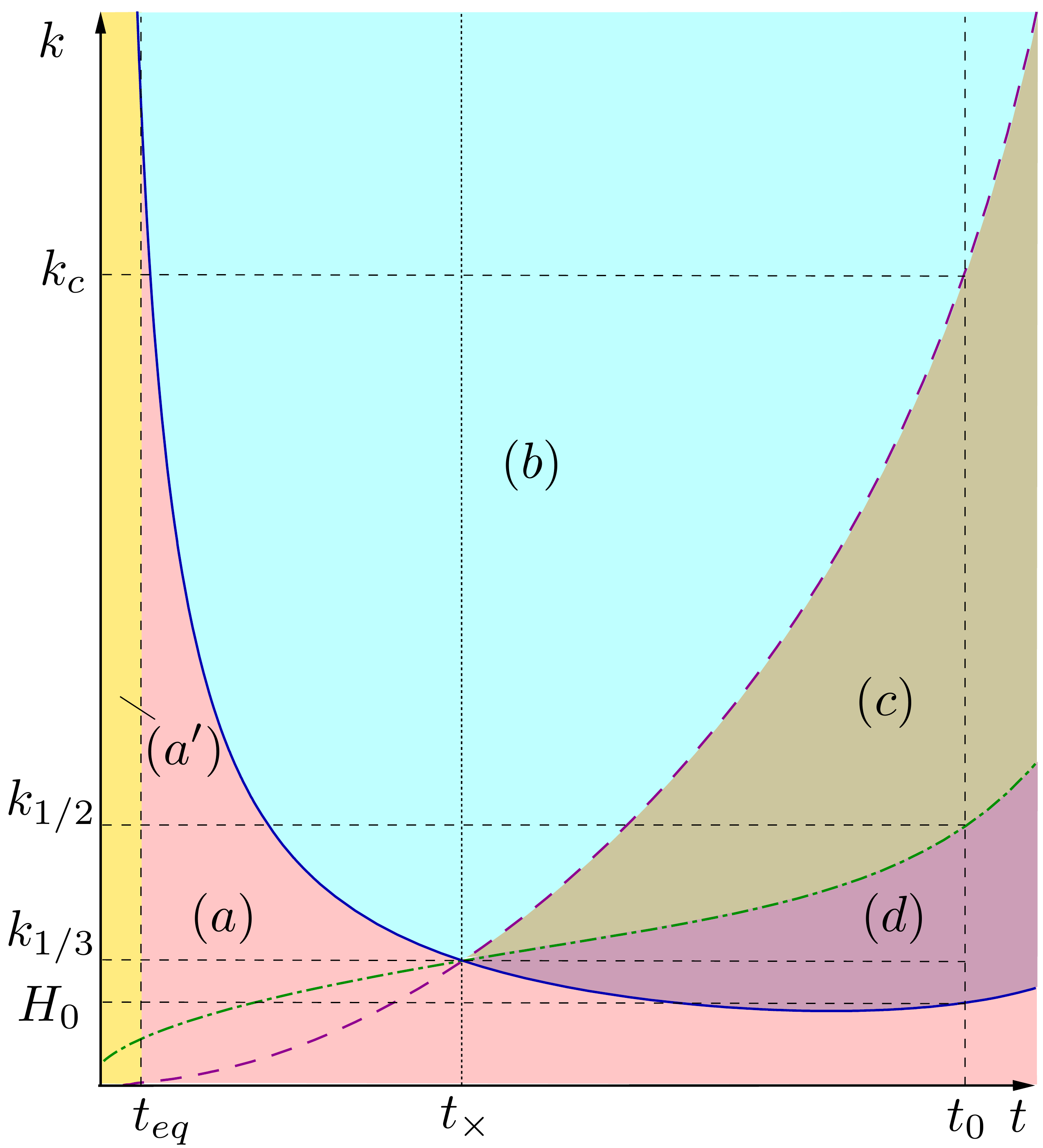}
\caption{Time dependence of different physical scales relevant for
  the dynamics of cosmological perturbations: $\HH$ (blue solid
  line), $ak_c$ (magenta dashed line), $\sqrt{\HH a k_c}$ (green
  dash-dotted line).  The time of the triple intersection of these
curves is denoted by $t_\times$. The evolution of perturbations proceeds
  differently in the regions {$ (a'), (a), (b), (c), (d)$}. The
  time of radiation-matter equality is denoted by $t_{eq}$; $t_0$ is the present time and  $H_0$
  is the present Hubble rate. The present scale factor $a(t_0)$ is set
  to $1$.
The beginning of current accelerated expansion roughly corresponds to the point where 
  $\HH$ starts growing.
The definition of $k_c$, $k_{1/2}$, $k_{1/3}$ is given in
  the text. 
\label{Fig:1}
}
\end{center}
\end{figure} 
One expects the behavior of a mode with a given comoving momentum $k$
to be different depending on the relation 
between\footnote{In principle, the comparison of the different scales
 should include a dependence on the propagation velocities
 $c_\Theta$ and $c_\chi$. For simplicity we will consider the
case where both quantities are close to one.} $k$ and various
scales that can be constructed out of the coefficients in
Eqs.~(\ref{khTetlin}). Clearly, two important scales are the Hubble
rate $\HH$ and the comoving value of the critical momentum
$ak_c$. Below we will encounter a third relevant scale, the geometric
mean of the first two, $\sqrt{\HH ak_c}$. The meaning of this latter
scale can be understood as follows. For modes with $k=\sqrt{\HH a
  k_c}$  the frequency (\ref{omega2}) of the low-frequency branch is
comparable to the expansion rate of the universe and therefore the
modes with this and smaller $k$
are significantly affected by the Hubble friction. 
The dependence of
the three scales on time is illustrated in Fig.~\ref{Fig:1}.
One distinguishes five different regimes of the mode evolution:
{}\\
${}\quad\quad${$ (a')$} superhorizon modes, $k<\HH$, radiation dominated universe;\\
${}\quad\quad${$(a)$} superhorizon modes, $k<\HH$, matter dominated universe;\\
${}\quad\quad${$(b)$} subhorizon modes with $\HH,\, ak_c<k$;\\
${}\quad\quad${$ (c)$} subhorizon modes with $\HH,\sqrt{\HH ak_c}<k<ak_c$;\\
${}\quad\quad${$(d)$} subhorizon modes with $\HH<k<\sqrt{\HH ak_c}$.

For the following discussion it is convenient to normalize $k$ to be
equal to the physical wave-number of the mode at present by setting
the today scale factor to unity, $a(t_0)=1$. 
It is also useful to express $k_c$ in terms of the present Hubble rate
$H_0$. The latter is given by the expression 
\[
H_0=c_\T \m^2\left(\frac{4\pi G_{cosm}}{3\Omega_{DE}}\right)^{1/2}\;,
\]
where $\Omega_{DE}\approx 0.75$ is the dark energy density fraction. Recalling
the definition (\ref{kc}) we obtain
\be
\label{kc1}
k_c=\frac{1}{c_\T}\left(\frac{6\Omega_{DE}[1+\b/2+3\l/2]}{\a}\right)^{1/2}H_0\;.
\ee
We see from  Fig.~\ref{Fig:1} that
depending on $k$ the different modes 
go through different regimes of evolution. Before
studying these regimes case by case, let us briefly 
outline   
their broad features.

The modes with $k>k_c$ go
only through the regimes {$ (a'), (a)$} and {$(b)$}. 
According to the analysis of Sec.~\ref{Sec:pertprev}, in these regimes
mixing between 
the  
$\chi$- and $\xi$-perturbations is weak and both modes have linear
dispersion relations with order one velocities. One expects that the
effect of these modes on the growth of cosmological perturbations is
at most of order $\a,\ \b,\ \l$, as it is the case for 
 pure Einstein-aether and khronon theories 
\cite{ArmendarizPicon:2010rs}.

The dynamics of the modes with $k_{1/2}<k<k_c$ are more interesting. Here 
\be
\label{k1/2}
k_{1/2}\equiv\sqrt{H_0 k_c}=\frac{1}{\sqrt{c_\T}}\left(\frac{6\Omega_{DE}[1+\b/2+3\l/2]}{\a}\right)^{1/4}H_0\;.
\ee 
These modes spend some time in the regime
{$ (c)$}. We are going to see that 
in this range there are non-trivial effects due to $\chi$ - $\xi$
mixing and the backreaction 
onto the  gravitational potentials is enhanced by a factor
$\a^{-1/2}$ compared to the short wave-length regime.
Note that for $\a$ not extremely small, $\a>10^{-8}$, the
momentum $k_{1/2}$ lies below the Hubble 
rate at the radiation-matter equality\footnote{\label{foot:11}This relation
  is obtained as follows. At the matter-dominated epoch $a\propto t^2$
and $\HH=2/t$. Hence, $\HH_{eq}=H_0(t_0/t_{eq})=H_0\sqrt{z_{eq}+1}$,
where $z_{eq}\sim 10^4$ is the corresponding red-shift.} 
$\HH_{eq}\sim 100 H_0$. Thus the modes with $k\sim k_{1/2}$
enter the horizon at the matter-dominated epoch.

The modes with $k_{1/3}<k<k_{1/2}$ pass through all five regimes 
{$ (a')$--$(d)$}. The momentum $k_{1/3}$ corresponds to the intersection of
the lines $\HH(t)$ and $a(t)k_c$, see Fig.~\ref{Fig:1}. To determine
it we note that for the interesting values of the parameters the
intersection happens at the epoch of matter domination. During this
epoch the scale factor is given by
\be
\label{amat}
a=A_{cm}t^2~,~~~~A_{cm}\equiv\frac{\Omega_{cm} H_0^2}{4}\;,
\ee 
where $\Omega_{cm}\approx 0.25$ is the matter density fraction. This
gives for the time of the intersection
\[
t_\times=\left(\frac{8}{\Omega_{cm}H_0^2k_c}\right)^{1/3},
\]
and hence,
\be
\label{k1/3}
k_{1/3}=\HH(t_\times)=(\Omega_{cm} H_0^2 k_c)^{1/3}
=\left(\frac{\Omega_{cm}}{c_\T}\right)^{1/3}\left(\frac{6\Omega_{DE}[1+\b/2+3\l/2]}{\a}\right)^{1/6}H_0\;.
\ee
For these modes we will also find a relative 
enhancement of the backreaction onto the gravitational potentials of order 
$\a^{-1/2}$.

Finally, the modes with $k<k_{1/3}$ go directly from the regime
{$(a)$} to the regime {$ (d)$} and do not have enough time to develop.
Thus the backreaction of these modes on the gravitational
potentials 
is expected to be $O(\a)$.
Note that for these modes to be
observable, their wave-length must be  inside the present horizon
size, $k>H_0$.  

We now confirm the statements made above by the detailed analysis of
the listed regimes.

\subsubsection*{Regime $(a')$: $k<\HH$ during radiation domination.}
As we are considering superhorizon modes, we can
neglect all terms with spatial Laplacians in
Eqs.~(\ref{khTetlin}). Some care is needed with the last term on the
l.h.s. of (\ref{xilin}) as it is the only term in this equation
containing the field $\chi$; we will check explicitly below that this
term is indeed small on the solutions. 

At the radiation-dominated stage the scale factor depends linearly on
time,
\be
\label{aRD}
a=A_\gamma t~,~~~~A_\gamma\equiv\sqrt{\Omega_\gamma}H_0\;,
\ee
where $\Omega_\gamma\sim 10^{-5}$ is
the radiation density fraction today. Substituting this and
(\ref{phiRDMD}) into Eqs.~(\ref{khTetlin}) we obtain
\bseq
\label{khTetlina'}
\begin{align}
\label{khlina'}
&\ddot\chi+\frac{2}{t}\dot\chi
+\bigg[\frac{2B}{t^2}+A_\gamma^2t^2c_\T^4k_c^2\bigg]\chi
+\frac{A_\gamma t\, c_\T^2k_c^2}{\m^2}\xi=\frac{1+B}{t}\phi_\gamma\;,\\
\label{xilina'}
&\ddot\xi+\frac{2}{t}\dot\xi
=-3A_\gamma\mu^2c_\T^2\phi_\gamma\;.
\end{align}
\eseq
This system has the solution:
\be
\label{chixia'}
\chi=\frac{\phi_\gamma t}{2}~,~~~\xi=-\frac{A_\gamma\m^2c_\T^2\phi_\gamma t^2}{2}\;.
\ee
It is straightforward to check that on this solution the
ratio of the term $\D\chi$ in
(\ref{xilin}) to the term $\ddot\xi$ is of order
$c_\T^2k^2t^2$ which is indeed small for superhorizon modes.

Note that despite the apparent growth of the fields $\chi$, $\xi$ in
(\ref{chixia'}), the physical perturbations remain subdominant with respect to the background. Indeed,
combining (\ref{chixia'}) with the expressions (\ref{sigmagauge}),
(\ref{attr}) for the background we obtain the total values of the
fields $\varphi$ and $\T$:
\begin{align}
&\varphi=t\,(1+\phi_\gamma/2)\;,\notag\\
&\T=-\frac{\m^2c_\T^2A_\gamma t^2}{2}(1+\phi_\gamma)\;.\notag
\end{align}
We observe that the relative corrections are small. The
solution (\ref{chixia'}) 
is equivalent to the time shift
\[
t\mapsto t\,(1+\phi_\gamma/2)\;,
\]
that corresponds precisely to the adiabatic mode at radiation
domination. 

Let us check whether
the solution \eqref{chixia'} is an attractor. To this end consider
 the solutions of
the corresponding homogeneous equations (i.e. Eqs. (\ref{khTetlina'}) with zero r.h.s.). Those
are particularly simple under the assumption that the second term in the square
brackets in (\ref{khlina'}) can be neglected. This is equivalent to
the assumption
\be
\label{ineq}
\HH\gg a k_c \;,
\ee 
which is always satisfied during radiation domination for reasonable
choices of parameters. Then the solution of the homogeneous equations
is
\bseq
\label{cxhom}
\begin{align}
\label{chom}
&\chi_{hom}=-\frac{A_\gamma c_\T^2k_c^2}{2(6+B)\m^2}\xi_0t^3
-\frac{A_\gamma c_\T^2k_c^2}{2(3+B)\m^2}Ct^2
+D_+t^{q_+}+D_-t^{q_-}\;,\\
\label{xhom}
&\xi_{hom}=\xi_0+\frac{C}{t}\;,
\end{align}
\eseq
where
\[
q_{\pm}=\frac{-1\pm\sqrt{1-8B}}{2}
\]
and $\xi_0$, $C$, $D_+$, $D_-$ are integration constants. 

Let us first
discuss the contributions proportional to $\xi_0$ and $C$. As long as
we concentrate on the field $\xi$, the role of these contributions
decreases with time compared to the adiabatic mode
(\ref{chixia'}). Moreover, the constant mode $\xi_0$ is irrelevant when
we are interested in the back-reaction of $\xi$ on the gravitational
potentials. Indeed, the only place where $\xi$ appears in the Einstein's
equations is Eq.~(\ref{Eins00}) and there it enters with a time
derivative. On the other hand, the terms proportional to $\xi_0$ and
$C$ in the homogeneous solution (\ref{chom}) for $\chi$ grow with time
faster than the adiabatic mode and may become
important at late times. Whether this happens or not depends on the
initial conditions for $\xi$ which set the values of $\xi_0$ and
$C$. A proper determination of these initial conditions would require
analysis of the evolution of the system at the epoch prior to
radiation-domination (i.e. at inflation and reheating) and also a
precise model for the origin of the field $\T$ (e.g. if it exists
during all the history of the universe, or appears as the result of a
phase transition at some particular epoch). These issues are outside
the scope of the present article. Nevertheless, let us give an
argument showing that $\xi_0$- and $C$-contributions into
$\chi$ are likely to be small. 
For concreteness we concentrate on the
$\xi_0$-contribution. Consider the ratio of this contribution to the
adiabatic mode (\ref{chixia'}). Omitting factors
of order one and using (\ref{Malpha}), (\ref{aRD}) 
it can be written in the form,
\be
\label{aratio}
A_\gamma t^2k_c\cdot\frac{\xi_0}{\phi_\gamma M_\a}
\sim \frac{a k_c}{\HH}\cdot\frac{\sqrt\epsilon\,\xi_0}{\sqrt\a H_{inf}}\;,
\ee   
where $H_{inf}$ is the Hubble rate at inflation and $\epsilon$ is the
inflationary slow-roll parameter. In passing to the second expression we
have used the standard formula 
\[
\phi_\gamma\sim \sqrt{\frac{G_{cosm}}{\epsilon}} H_{inf},
\]
for the amplitude of the primordial perturbations in the potentially
driven slow-roll inflation. 
The first factor in (\ref{aratio}) is small due to the inequality
(\ref{ineq}). Let us estimate the second factor. Assuming that the
fluctuations of the $\xi$-field are generated at the inflationary
epoch, we have the estimate $\xi_0\lesssim H_{inf}$. We will be
interested in the values of $\a$
in the range from $10^{-4}$ to
$10^{-2}$. Finally, recalling that the measurements of the tilt of the
primordial spectrum give an upper bound $\epsilon\lesssim 10^{-2}$, we
obtain that the second factor in (\ref{aratio}) can be at most $\lesssim10$.   
Overall, we conclude that the product (\ref{aratio}) is small in general.
A similar reasoning
applies to the contribution in (\ref{chom}) that is proportional to
$C$. We will neglect these contributions in what follows. 

Consider now the last two terms in (\ref{chom}). The term proportional
to $D_-$ always decays with time and can be safely neglected. However,
the $D_+$ term can either decay or grow depending on the sign of the
parameter $B$. Moreover, for $B<-1$ this term grows faster than the
adiabatic contribution (\ref{chixia'}). We have already encountered this
growing regime in Sec.~\ref{Sec:pertprev}, where it was
related to the possibility of having negative khronon mass
squared. Below we will only consider the
case $B>-1$ where the leading solution
is given by the adiabatic mode (\ref{chixia'}). 

\subsubsection*{Regime $(a)$: $k<\HH$ during matter domination.}
During matter domination the scale factor
is given by the formula (\ref{amat}). Substituting this into
Eqs.~(\ref{khTetlin}) and neglecting the terms with spatial Laplacians
we obtain
\bseq
\label{khTetlina}
\begin{align}
\label{khlina}
&\ddot\chi+\frac{4}{t}\dot\chi
+\bigg[\frac{2(1+3B)}{t^2}+A_{cm}^2t^4c_\T^4k_c^2\bigg]\chi
+\frac{A_{cm}t^2 c_\T^2k_c^2}{\m^2}\xi=\frac{2(1+B)}{t}\phi_{cm}\;,\\
\label{xilina}
&\ddot\xi+\frac{4}{t}\dot\xi
=-6A_{cm}t\mu^2c_\T^2\phi_{cm}\;.
\end{align}
\eseq
The adiabatic mode is also a solution and has the form,
\be
\label{chixia}
\chi=\frac{\phi_{cm}t}{3}~,~~~~\xi=-\frac{A_{cm}\m^2c_\T^2\phi_{cm}t^3}{3}\;.
\ee 
It corresponds to the time shift
\[
t\mapsto t(1+\phi_{cm}/3)\;.
\]
Similarly to the previous case one can check that the term proportional to $\Delta \chi$ on 
(\ref{xilin}) can be safely ignored. 
Also, as in the previous case,
one can argue that 
the solutions of the homogeneous equations 
are irrelevant, apart from one, possibly
growing, mode. This mode appears in the regime (\ref{ineq}) and has
the form
\[
\chi_{hom}\propto t^{r_+}~,~~~~r_+=\frac{-3+\sqrt{1-24B}}{2}\;.
\]
It grows faster than the adiabatic mode if $B<-1$. This
growth is cut off once the mode enters inside the horizon or the 
inequality (\ref{ineq}) is violated. Below we restrict 
to the stable case $B>-1$.

\subsubsection*{Regime $(b)$: subhorizon modes with $ak_c<k$.}

We now turn to the evolution of the subhorizon
perturbations. We shall concentrate on the modes that enter inside
the horizon at the matter-dominated stage. 
Neglecting the $\HH$-dependent terms in brackets 
in Eq.~(\ref{khlin}) and substituting (\ref{phiRDMD}) on the r.h.s. 
we obtain,
\bseq
\label{khTetlinb}
\begin{align}
\label{khlinb}
&\ddot\chi+\frac{4}{t}\dot\chi
+\big[c_\chi^2k^2+A_{cm}^2t^4c_\T^4k_c^2\big]\chi
+\frac{A_{cm}t^2c_\T^2k_c^2}{\m^2}\xi=\frac{2(1+B)}{t}\phi_{cm}\;,\\
\label{xilinb}
&\ddot\xi+\frac{4}{t}\dot\xi+c_\T^2k^2\xi+
A_{cm}t^2\m^2c_\T^4k^2\chi
=-6A_{cm}t\mu^2c_\T^2\phi_{cm}\;.
\end{align}
\eseq
Deep inside the regime under study, $k\gg ak_c$, the previous equations have the approximate solution:
\be
\label{chixib}
\chi=\frac{2(1+B)\phi_{cm}}{c_\chi^2k^2t}~,~~~~
\xi=-2\bigg(3+\frac{c_\T^2}{c_\chi^2}(1+B)\bigg)
\frac{A_{cm}\m^2\phi_{cm} t}{k^2}\;.
\ee
This solution 
decays with time (in the case of the $\xi$-perturbation
it must be compared with the behavior $\bar\T\propto t^3$ of the
background). On top of (\ref{chixib}) the fields $\chi$, $\xi$ exhibit
oscillations with  frequencies $\omega_\chi $, $\omega_\xi$
(Eqs.~(\ref{displarge})) and 
 amplitudes decaying as $1/t^2$.
These oscillations
represent small corrections to the approximate solution (\ref{chixib}) which 
disappear at large times.

\subsubsection*{Regime $(c)$: subhorizon modes with $\sqrt{\HH ak_c}<k<ak_c$.}

According to the results of
Sec.~\ref{Sec:pertprev}, in this regime one of the branches of
$\chi$--$\xi$ perturbations has a frequency gap  of order $k_c$, see Eq.~\eqref{omega1}. As long as we consider frequencies 
smaller than the gap, this mode can be integrated
out. To obtain the equations for the remaining mode we
use the following trick. 
First, we eliminate the explicit dependence of the equations on the
parameter $\m$ by introducing 
\[
\tilde \xi =\xi/\m^2\;.
\]
The second step is to take the limit
\bseq
\be
\label{kcinf}
k_c\to \infty\;.
\ee
The physical meaning of this limit is that we focus on the modes with
frequencies $\omega$ much smaller than the gap (cf. (\ref{omega2})). At the same time we
must be careful to keep the frequency of these modes
non-zero. From Eq.~(\ref{omega2}) we read out $\omega_-\propto
k^2/k_c$. Thus to keep $\omega_-$ fixed (\ref{kcinf}) must be
accompanied by 
\be
\label{kinf}
k\to\infty~,~~~~k\propto \sqrt{k_c}\;.
\ee
\eseq
The rest of the quantities in the equations must remain 
finite\footnote{An alternative derivation of this limit and
  Eqs.~(\ref{khTetlinc})
  involves diagonalization of
  the quadratic action (\ref{S2FRW}).}.
In this way Eq.~(\ref{khlinb}) reduces
to
\bseq
\label{khTetlinc}
\be
\label{khlinc}
\bigg[\frac{c_\chi^2 k^2}{A_{cm}t^2c_\T^2k_c^2}+A_{cm}t^2c_\T^2\bigg]\chi
+\tilde\xi=0\;,
\ee
where we have kept the main subleading contribution represented by the
first term in the brackets. Substituting this into (\ref{xilinb}) we
obtain,
\be
\label{xilinc}
\ddot{\tilde\xi}+\frac{4}{t}\dot{\tilde\xi}
+\frac{c_\chi^2k^4}{A_{cm}^2t^4c_\T^2k_c^2}\tilde\xi=
-6A_{cm} t c_\T^2\phi_{cm}\;.
\ee
\eseq
In the regime {$ (c)$} that we are considering now, the leading
solution is obtained by neglecting the terms with time derivatives on
the l.h.s. of (\ref{xilinc}). This gives,
\be
\label{chixic}
\chi=\frac{6A_{cm}^2c_\T^2k_c^2t^3\phi_{cm}}{c_\chi^2k^4}~,~~~
\xi=-\frac{6A_{cm}^3\m^2c_\T^4k_c^2\,t^5\phi_{cm}}{c_\chi^2k^4}\;,
\ee 
where the expression for $\chi$ is obtained using
(\ref{khlinc}). Clearly, these perturbations grow with time. As 
anticipated at the end of  Sec.~\ref{Sec:pertprev} this growth results
from the low propagation speed of these modes.

On top of the previous solution there is an oscillatory mode that
corresponds to the 
solution of the homogeneous part of (\ref{xilinc}),
\[
\xi_{hom}\propto\frac{1}{t}\cdot
\exp\bigg(\pm i\frac{c_\chi k^2}{A_{cm}c_\T k_ct}\bigg)\;.
\]
However, it rapidly decays compared to (\ref{chixic}) and hence is
irrelevant.

\subsubsection*{Regime $(d)$: subhorizon modes with $k<\sqrt{\HH ak_c}$.}

In this case the frequency of the modes is
lower than the Hubble rate and we can neglect the third term on the
l.h.s. of (\ref{xilinc}). Then the perturbations are frozen and we
obtain the same adiabatic solution (\ref{chixia}) as in the superhorizon
case. It is important to stress though, that the modes we are
discussing now have wave-lengths shorter than the Hubble size and
are therefore observable. 

\subsection{Corrections to the metric perturbations}

According to the iterative procedure described at the beginning of
the current section, we shall substitute
the expressions for the $\chi$ and $\xi$ perturbations obtained above into the
Einstein's Eqs.~(\ref{Einslin}). This will give us the corrections to
the observable quantities due to the presence of these fields.
 It is convenient to concentrate on
Eq.~(\ref{Einsij1}). Substituting $\phi$ in terms of $\psi$ and $\chi$
from
(\ref{Einsij2}) we obtain
\be
\label{Einsmode2}
\begin{split}
\ddot\psi&+3\HH \dot\psi+(2\dot\HH+\HH^2)\psi
-\frac{a^2\delta p_{[\g]}}{(2+\a B)M_0^2}
=\\
&-\frac{\b+\l}{2+\a B}(\D\dot\chi+2\HH\D\chi)-\b\big[\HH\ddot\chi+(2\dot\HH+3\HH^2)\dot\chi
+(6\dot\HH\HH+2\HH^3)\chi\big]\;.
\end{split}
\ee 
The r.h.s. represents the source for the correction $\psi^{(1)}$ to
the standard adiabatic behavior (\ref{phiRDMD}) 
 of $\psi$. Let us compare the two
terms on the r.h.s. For superhorizon modes the second term
dominates. Clearly, it gives a contribution of order $O(\b)$ to
$\psi^{(1)}$. On the other hand, for subhorizon modes the first term
becomes dominant. Indeed, it is
enhanced compared to the second term by the ratio $(k/\HH)^2$. 
Below we focus on the contribution of this term into $\psi^{(1)}$ and
neglect the second term on the r.h.s. of (\ref{Einsmode2}). 

Assuming as before that the relevant modes enter the horizon at
the matter-dominated epoch, Eq.~(\ref{Einsmode2}) is further
simplified to
\[
\ddot\psi+\frac{6}{t}\dot\psi=\frac{(\b+\l)k^2}{2}
\bigg(\dot\chi+\frac{4}{t}\chi\bigg)\;.
\]
This yields for $\psi^{(1)}$ the expression
\be
\label{psi1}
\psi^{(1)}(t)=\frac{(\b+\l)k^2}{2}\int_{t_*}^t \frac{\di t'}{{t'}^6}
\int_{t_*}^{t'}\di t'' {t''}^6\bigg(\dot\chi(t'')+\frac{4}{t''}
\chi(t'')\bigg)\;,
\ee
where $t_*$ is the time of horizon crossing
for the mode with given $k$. At this time, the adiabatic initial conditions
imply $\psi^{(1)}(t_*)\approx 0$.

Consider first the modes with $k_c<k$. As it is clear from
Fig.~\ref{Fig:1}, these modes spend all subhorizon evolution in the
regime {$ (b)$}, where the $\chi$-field is given by
(\ref{chixib}). Inserting this into (\ref{psi1}) gives,
\be
\label{psi1short}
\psi^{(1)}\sim (\b+\l)\cdot\frac{3(1+B)}{5c_\chi^2}
\phi_{cm}\ln\frac{t}{t_*}\;.
\ee
We see that for these short modes the correction to the
gravitational potential remains parametrically small,
i.e. $O(\b+\l)$. Notice, however, that the numerical value of the
logarithmic factor in (\ref{psi1short}) can be quite large (of order
10), which can lead to a sizable effect if $\b,\ \l$ are not too small.

Next we turn to the modes with $k_{1/2}<k<k_c$, see
Fig.~\ref{Fig:1}. Due to fast growth of these modes in the region
$(c)$, the integrals in (\ref{psi1}) are saturated at the upper end of
the integration domain. Using (\ref{chixic})
we get an estimate for the value of $\psi^{(1)}$ at
present\footnote{Strictly speaking, the formula (\ref{psi1}) is
  applicable only during matter-domination. However, for the reasons
  discussed at the beginning of this section, 
we neglect possible corrections due to the
  start of the accelerated cosmological expansion.}:
\[
\psi^{(1)}(t_0)\approx \frac{7}{12}(\b+\l)
\frac{c_\T^2k_c^2}{c_\chi^2k^2}\phi_{cm}\;.
\]
At $k\approx k_{1/2}$ we obtain
\be
\label{psimax}
\psi^{(1)}(t_0)\big|_{k=k_{1/2}}\approx\frac{7\sqrt{6\Omega_{DE}}c_\T}{12
c_\chi^2}\cdot\frac{\b+\l}{\sqrt\a}\phi_{cm}\;,
\ee
where we have used Eqs.~(\ref{kc1}), (\ref{k1/2}). We clearly see the
parametric 
enhancement of the signal to the level\footnote{Recall that we assume
  the parameters $\a,\ \b,\ \l$ to be of the same order.} $O(\sqrt\a)$.

Finally, we consider the modes with $H_0<k<k_{1/2}$. Again, the
integral (\ref{psi1}) is saturated at the upper end corresponding to
the regime {$ (d)$}. Using the appropriate formula for $\chi$,
Eq.~(\ref{chixia}), we obtain
\[
\psi^{(1)}(t_0)\approx \frac{5}{84}(\b+\l)(kt_0)^2\phi_{cm}
\approx\frac{5}{21}(\b+\l)\bigg(\frac{k}{H_0}\bigg)^2\phi_{cm}\;,
\]
where in passing to the last expression we have used $H_0\approx
2/t_0$. This expression has a maximum at the highest momenta of the
interval, 
$k\sim k_{1/2}$. Using (\ref{k1/2}) one readily verifies that, up
to a numerical 
factor of order one, this expression matches at $k\sim k_{1/2}$  with the
estimate~(\ref{psimax}). 

To sum up, the analytical study of this section shows that 
the correction $\psi^{(1)}_k$ to the present-day amplitude
of the gravitational potential has a maximum
of order $\sqrt\a\phi_{cm}$ at $k\sim k_{1/2}$ with the falloff
$\psi^{(1)}\propto k^2$, $\psi^{(1)}\propto k^{-2}$ on the two sides,
and a logarithmic tale $\psi^{(1)}\sim (\b+\l)\phi_{cm}\ln{k}$ extending
towards large momenta $k>k_c$.  This will be verified by the 
numerical results in the next section.

\section{Cosmological perturbations: numerical results}
\label{Sec:pertnum}

\begin{figure}[!htb]
\begin{center}
\includegraphics[width=0.45\textwidth]{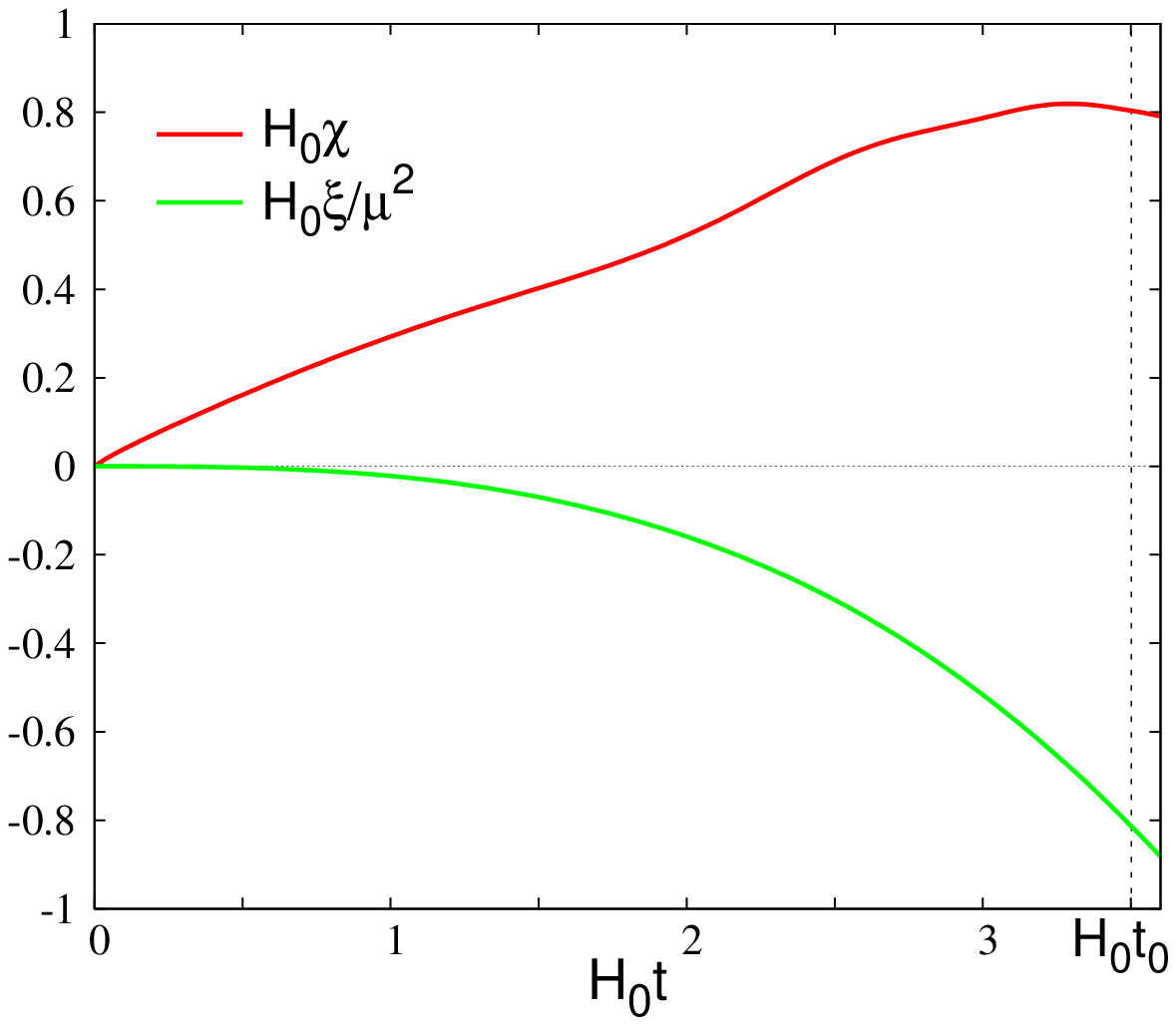}
\includegraphics[width=0.54\textwidth]{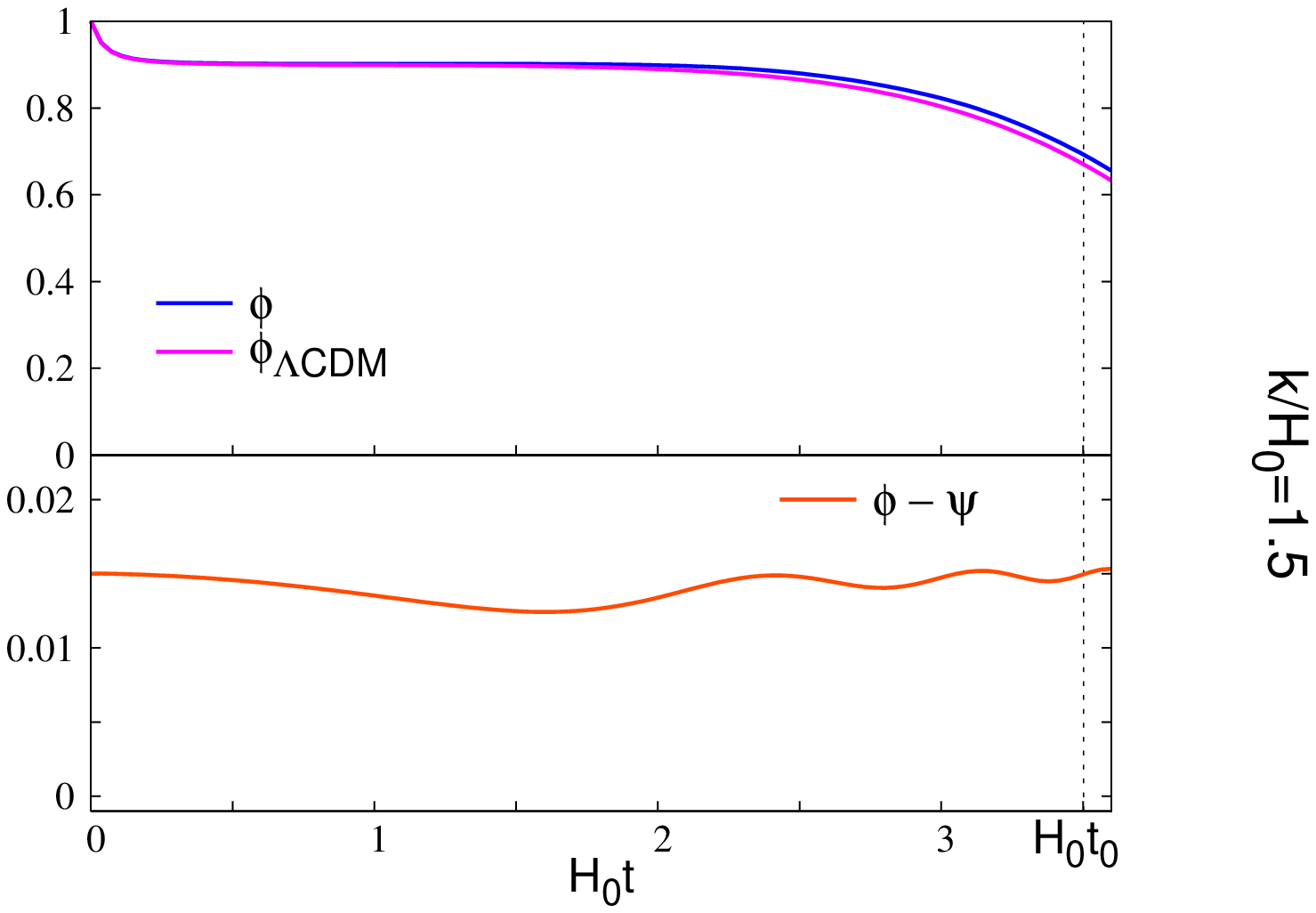}\\
\includegraphics[width=0.45\textwidth]{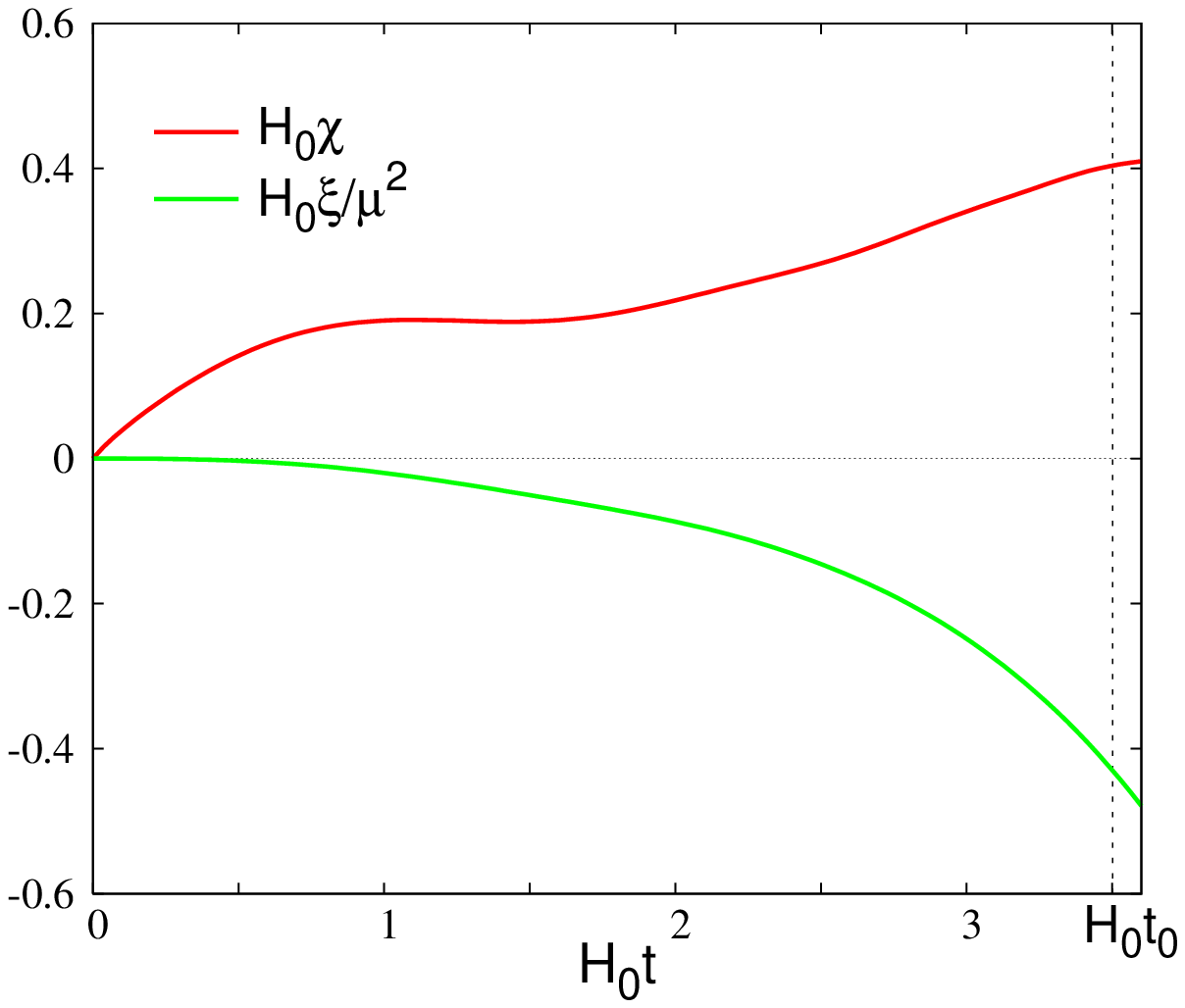}
\includegraphics[width=0.54\textwidth]{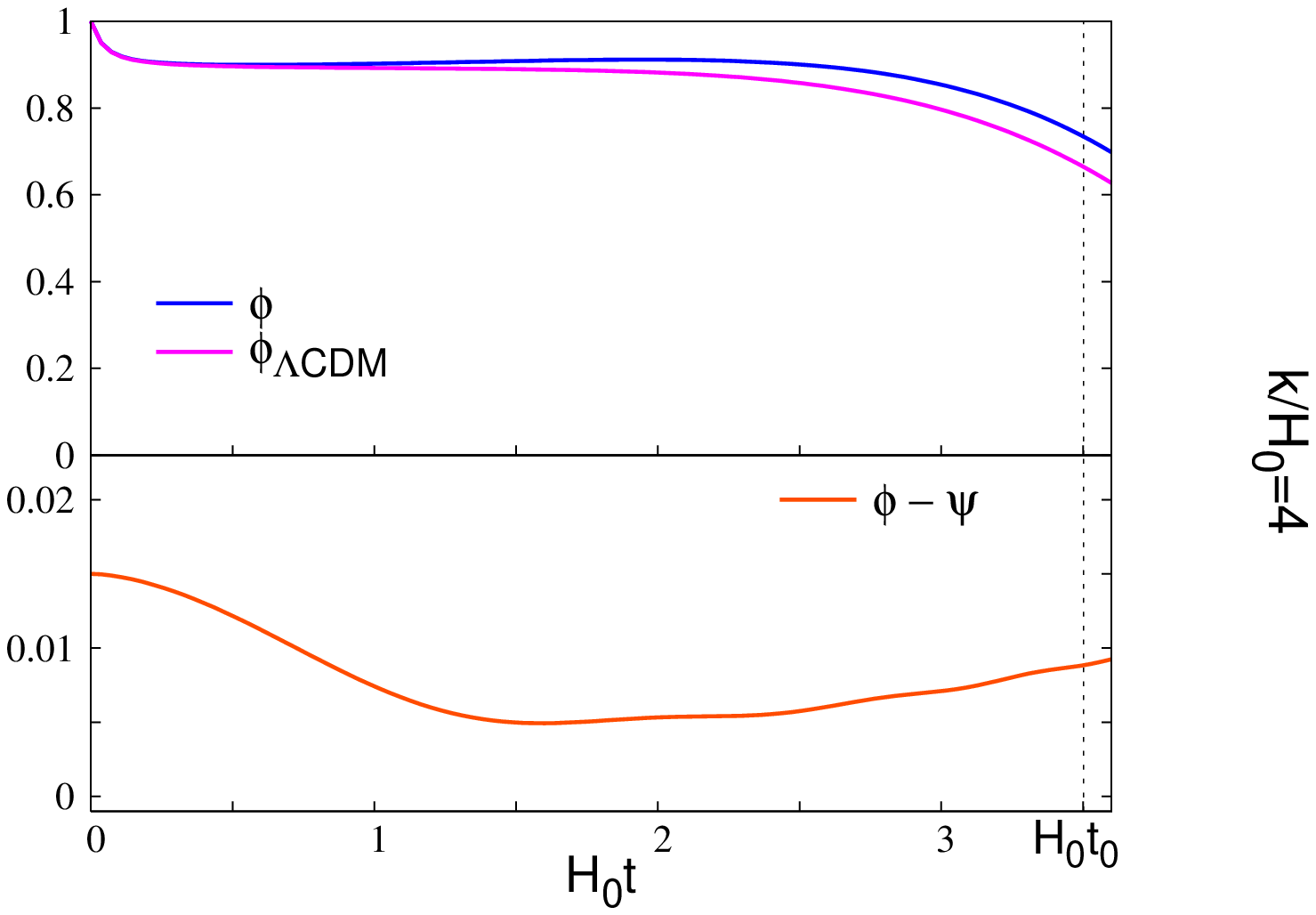}
\caption{Solution of the system (\ref{Einslin}), (\ref{khTetlin}),
  (\ref{matter}) for two values of the momentum: $k=1.5 H_0$
  (upper panels), $k=4 H_0$ (lower panels), where $H_0$ is the present
  Hubble constant. Left panels show the dependence of the dark energy perturbations
  $\chi$, $\xi$ on conformal time $t$. Right panels show the
  perturbations of the gravitational potential $\phi$ and the
  difference $(\phi-\psi)$ between the two gravitational potentials. For
  comparison we also present the dependence of the gravitational
  potential in the standard $\Lambda$CDM cosmology. 
The parameters of the
  model are $\a=0.02$, $\b=0.01$, $\l=0.01$. For these values $k_c=15.15
  H_0$, $k_{1/2}=3.89 H_0$, $k_{1/3}=1.56 H_0$. Present time
  corresponds to $t_0= 3.5 H_0^{-1}$.
\label{Fig:2}
}
\end{center}
\end{figure}
\begin{figure}[!tb]
\begin{center}
\includegraphics[width=0.45\textwidth]{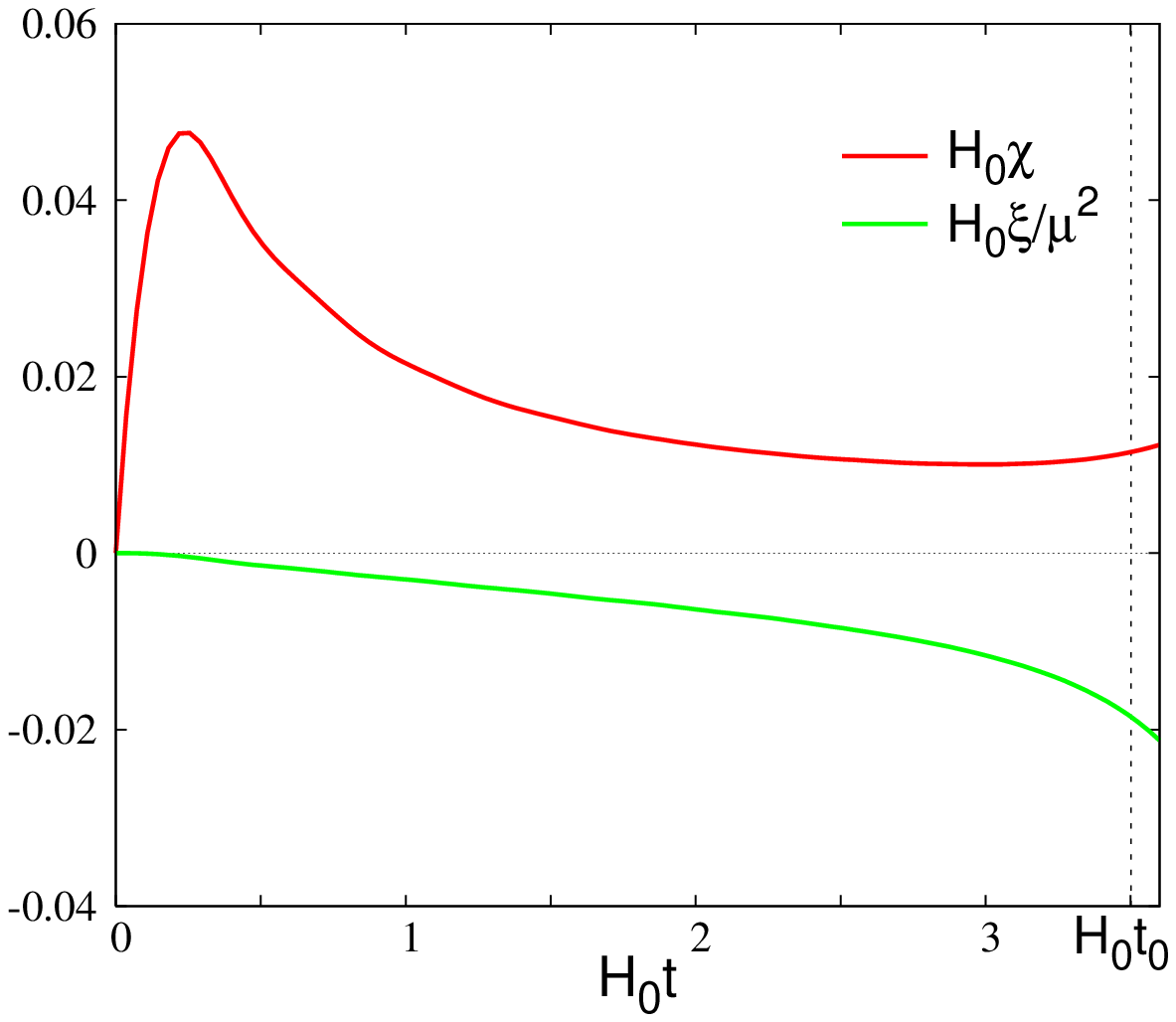}
\includegraphics[width=0.54\textwidth]{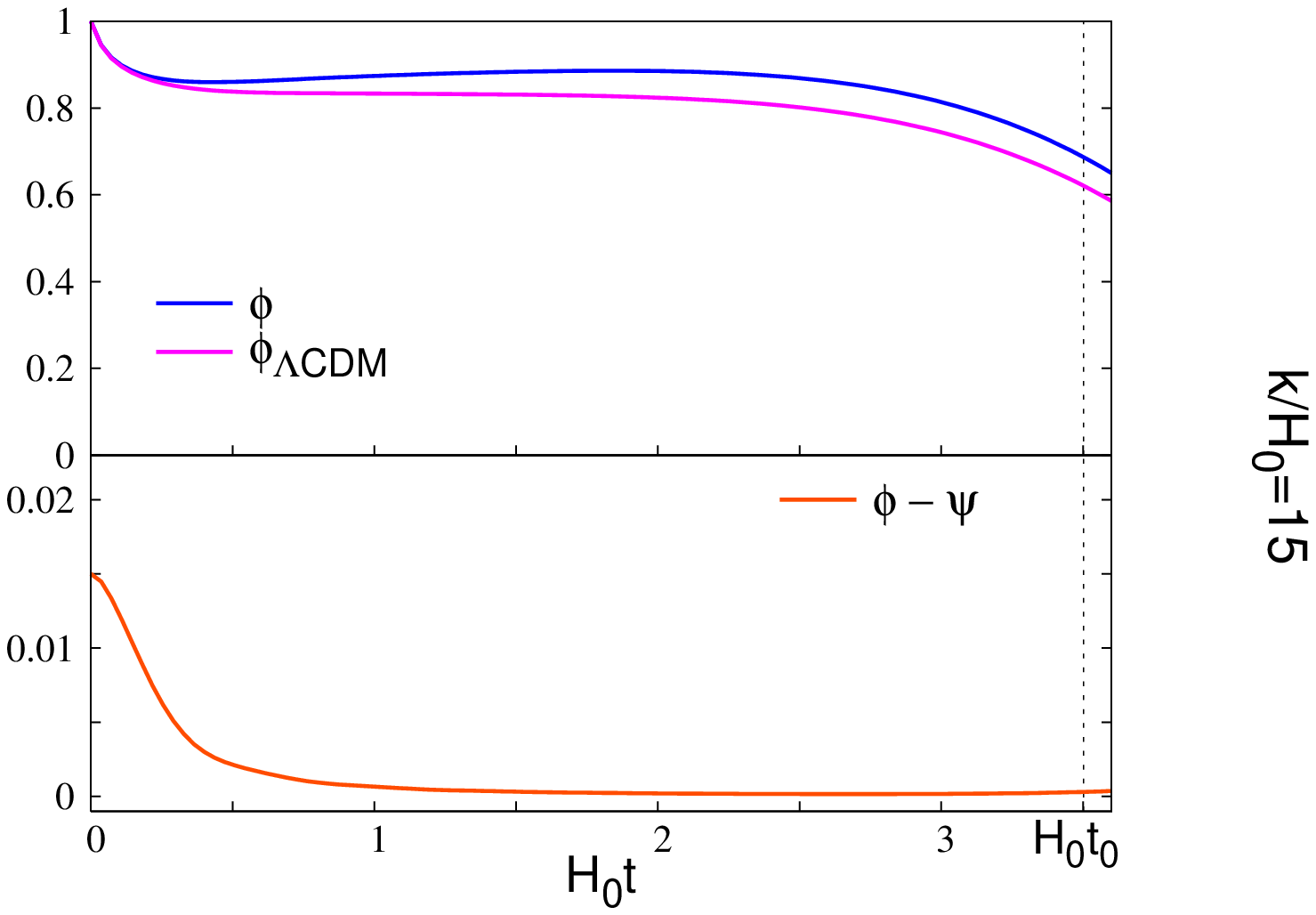}\\
\includegraphics[width=0.45\textwidth]{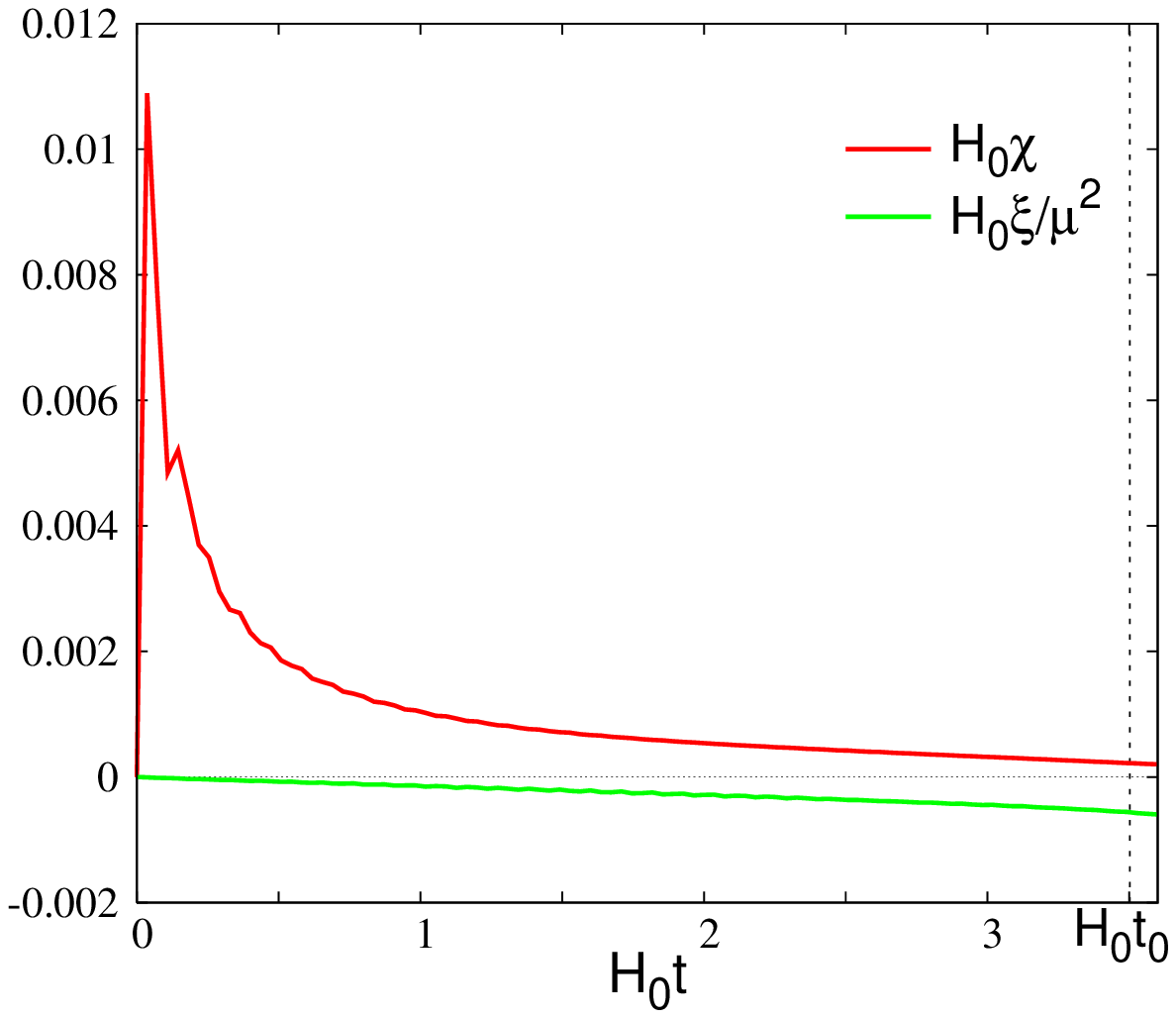}
\includegraphics[width=0.54\textwidth]{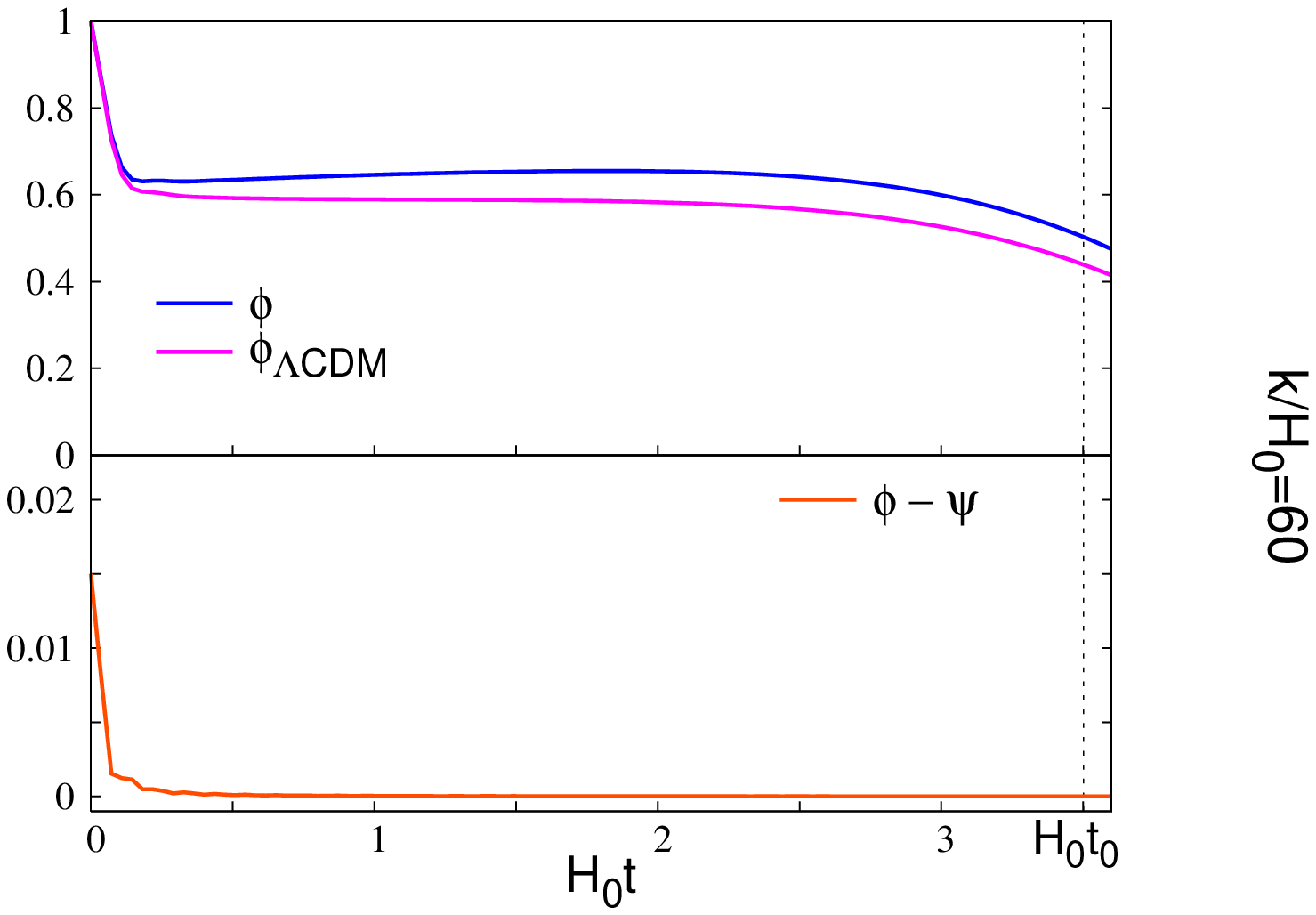}
\caption{The same as Fig.~\ref{Fig:2} for $k=15 H_0$ (upper panels),
  $k=60 H_0$ (lower panels). 
\label{Fig:3}
}
\end{center}
\end{figure} 

In this section we report the results of numerical integration of
Eqs.~(\ref{Einslin}), (\ref{khTetlin}), (\ref{matter}). We consider
the density fractions of radiation and cold matter today
\[
\Omega_{\gamma}=5\cdot 10^{-5}~,~~~\Omega_{cm}=0.25\;.
\]
Correspondingly, the dark energy fraction is
\[
\Omega_{DE}=0.75\;.
\]
Let us emphasize again that for simplicity we neglect all effects related to baryon-photon
coupling and merely include baryons into the cold matter
fraction. Similarly, we neglect the neutrino masses. Still,
within these simplifications we refer to the cosmological model with CC
and without 
$\xi$ and $\chi$ perturbations as $\Lambda$CDM.
Details of the numerical procedure are presented in the
Appendix~\ref{App:A}. Here we summarize the results.

The time dependence of the fields $\chi$, $\xi$, 
the gravitational potential $\phi$ and the difference
$(\phi-\psi)$ is shown in Figs.~\ref{Fig:2} and \ref{Fig:3}
for several values of the
momentum $k$. The parameters of the model are taken to be
\[
\a=0.02~,~~~\b=0.01~,~~~\l=0.01~,~~~c_\T=1\;.
\]
This choice satisfies the PPN\footnote{Recall that $\a=2\b$ case
  avoids the PPN bounds.}, BBN and gravitational radiation
constraints \cite{Blas:2010hb,BlasSanc}.
The initial conditions correspond to the adiabatic mode, and, for
illustration purposes, the 
initial value of $\phi$ is normalized to 1 for every momentum. We also
show for comparison the dependence of the Newton potential on time 
for the standard
$\Lambda$CDM case (where $\psi=\phi$ within our approximation).

\begin{figure}[!tb]
\begin{center}
\includegraphics[width=0.6\textwidth]{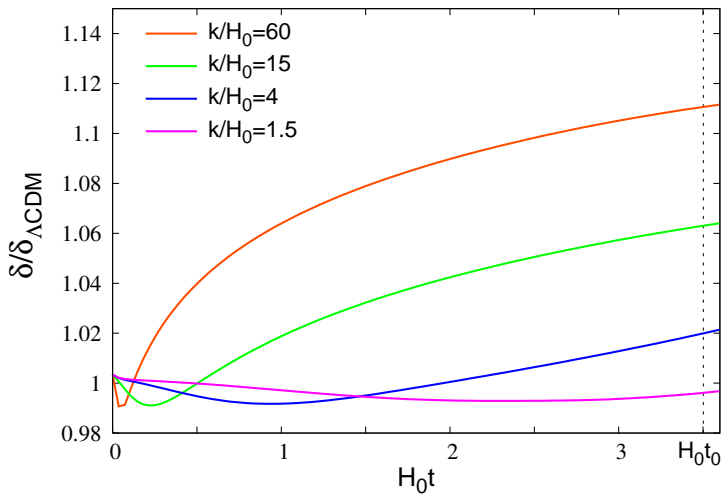}
\caption{The ratio of density contrasts in the $\T$CDM and 
$\Lambda$CDM models versus time for several values of the
  mode momentum $k$. The values of parameters are the same
  as for Fig.~\ref{Fig:2}.  
\label{Fig:4}
}
\end{center}
\end{figure} 

We see that perturbations of the gravitational potential are enhanced
at late times compared to the $\Lambda$CDM case. The relative
magnitude of enhancement is of order $10\%$ which agrees with the
analytic estimate $\sqrt\a\approx 0.14$ derived in the previous
section. On the other 
hand, the difference between the two gravitational potentials 
$(\phi-\psi)$ which is initially of order $10^{-2}$ rapidly decreases
once the mode enters into the horizon, and as a consequence is at present
negligible for $k$ larger than $k_{1/2}$. Note that the overall
amplitude $10^{-2}$ for long wavelength modes 
agrees with the estimates $(\phi-\psi)\sim\a=0.02$.

One expects that the enhancement in the gravitational perturbations
will lead to the increase of structure growth rate. To illustrate
this point we plot in Fig.~\ref{Fig:4} the ratio
$\delta/\delta_{\Lambda CDM}$, where $\delta$ and $\delta_{\Lambda
  CDM}$ are the cold matter density contrasts in the case of the
present model and $\Lambda$CDM respectively,
\be
\label{eq:contrastcm}
\delta\equiv\frac{\delta \r_{[cm]}}{\bar \r_{[cm]}}.
\ee
As expected we observe the increase in the growth of structure at
recent times. The relative effect is stronger for shorter modes and
can be as large as $11\%$ for our choice of parameters. 

\begin{figure}[!tb]
\includegraphics[width=0.49\textwidth]{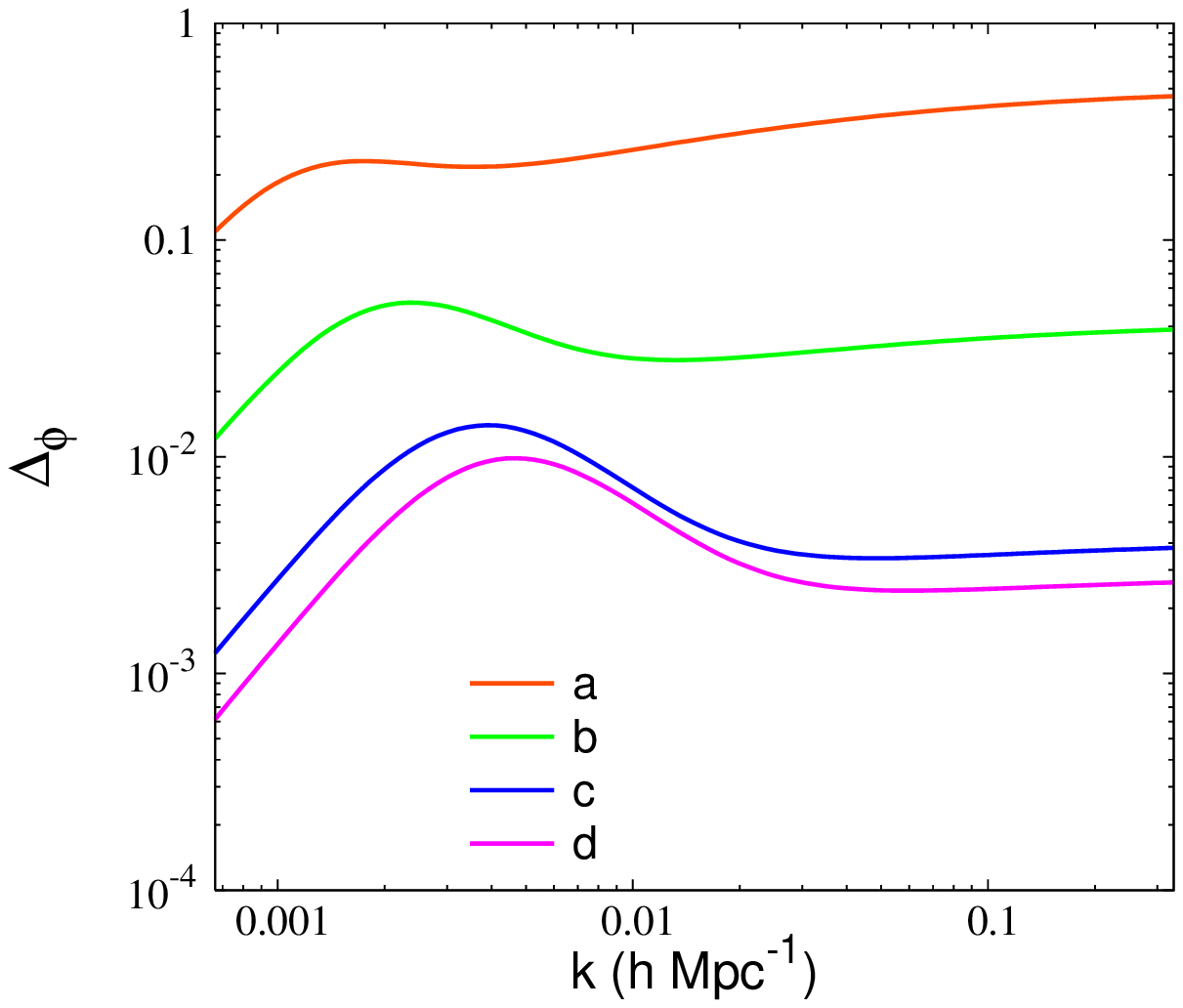}
\hfill
\includegraphics[width=0.49\textwidth]{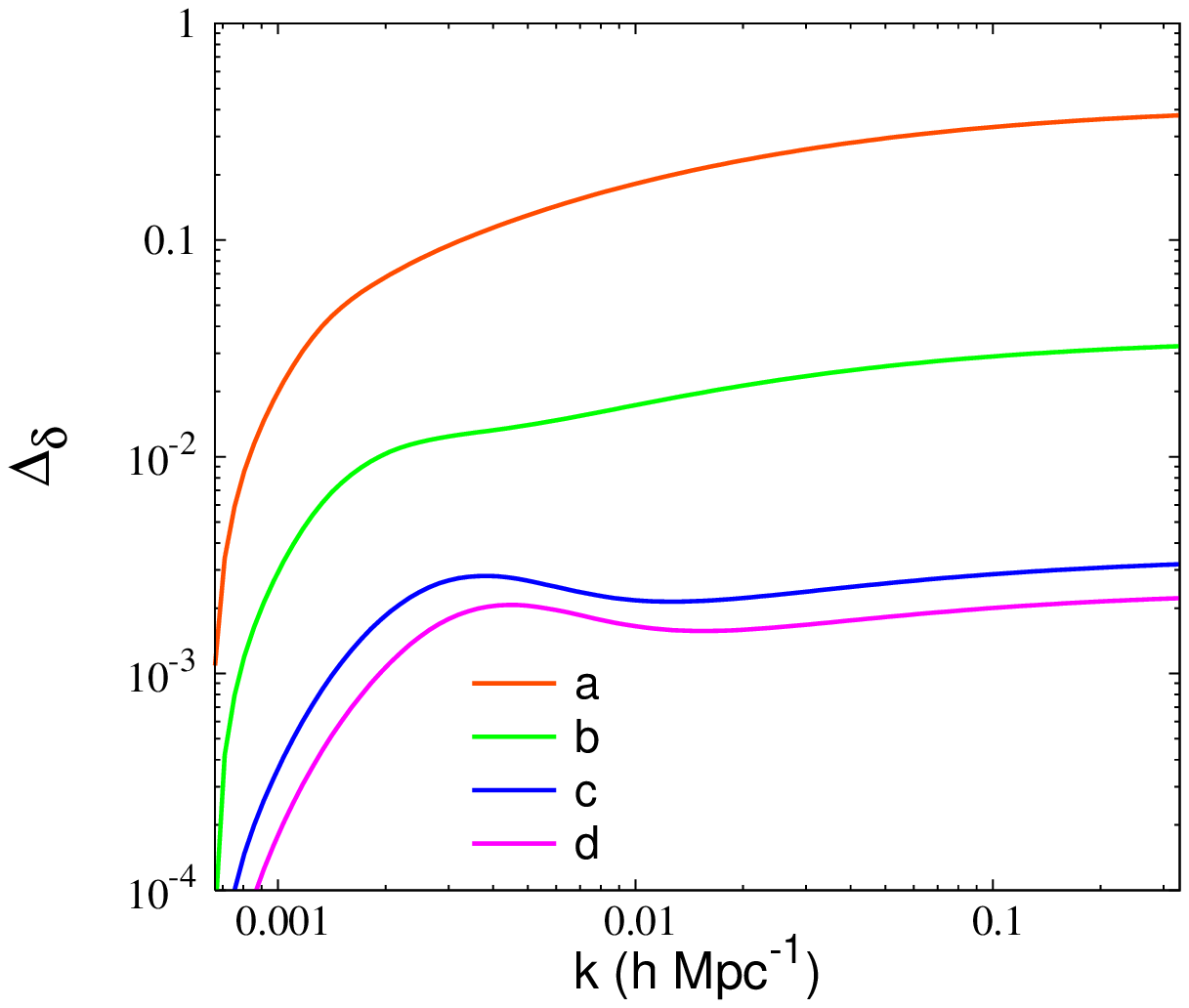}
\caption{Comparison between the power spectra for the gravitational
  potential (left panel) and the matter density contrast (right panel)
  in the $\T$CDM and $\Lambda$CDM
  cosmologies. The curves correspond to the values of 
the model parameters listed in Table~\ref{tab:1}.
\label{Fig:5}
}
\end{figure}

The next plots show the comparison between the power spectra of
perturbations in the $\T$CDM and 
$\Lambda$CDM cosmologies evaluated at the present moment of time. 
Fig.~\ref{Fig:5} shows the relative
differences of the power spectra for the gravitational potential $\phi$ and the
cold matter density contrast $\delta$,
\[
\Delta_\phi(k)\equiv\frac{P_\phi(k)}{P_{\phi_{\Lambda CDM}}(k)}-1,\quad
\Delta_{\delta}(k)\equiv\frac{P_{\delta}(k)}{P_{\delta_{\Lambda CDM}}(k)}-1.
\] 
These differences were computed for several values
of the model parameters that are listed in Table~\ref{tab:1}. 
We also present the corresponding values of the momenta $k_c$ and
$k_{1/2}$. 
Note that, as expected from the results of the previous section, 
the functions $\Delta_i(k)$ peak around $k\approx k_{1/2}$ 
(different for the different sets of parameters) 
with a tail extending to larger momenta. 
In agreement with the analytical results, the value of the functions $\Delta_i$ 
at the peak scales
approximately as $\sqrt\a$ while the tails are proportional to $\a$.  
\begin{table}
\begin{center}
\begin{tabular}{|c|c|c|c|c|c|}
\hline
 & $\a$ & $\b$ & $\l$ & $k_c$ (h Mpc$^{-1}$) & $k_{1/2}$ (h Mpc$^{-1}$) \\\hline
a & $2\cdot 10^{-2}$ & $10^{-2}$ & $10^{-2}$ & $5.1\cdot 10^{-3}$ &
$1.3\cdot 10^{-3}$\\\hline
b & $2\cdot 10^{-3}$ & $10^{-3}$ & $10^{-3}$ & $1.6\cdot 10^{-2}$ & $2.3\cdot 10^{-3}$\\\hline
c & $2\cdot 10^{-4}$ & $10^{-4}$ & $10^{-4}$ & $5.0\cdot 10^{-2}$ & $4.1\cdot 10^{-3}$\\\hline
d & $10^{-4}$ & $0$ & $10^{-4}$ & $7.1\cdot 10^{-2}$ & $4.9\cdot 10^{-3}$\\\hline
\end{tabular}
\end{center}
\caption{Model parameters corresponding to the curves in
  Fig.~\ref{Fig:5} and Fig.~\ref{Fig:6} and the respective
  values of momenta $k_c$, $k_{1/2}$.}
\label{tab:1}
\end{table}

\begin{figure}[!tb]
\begin{center}
\includegraphics[width=0.6\textwidth]{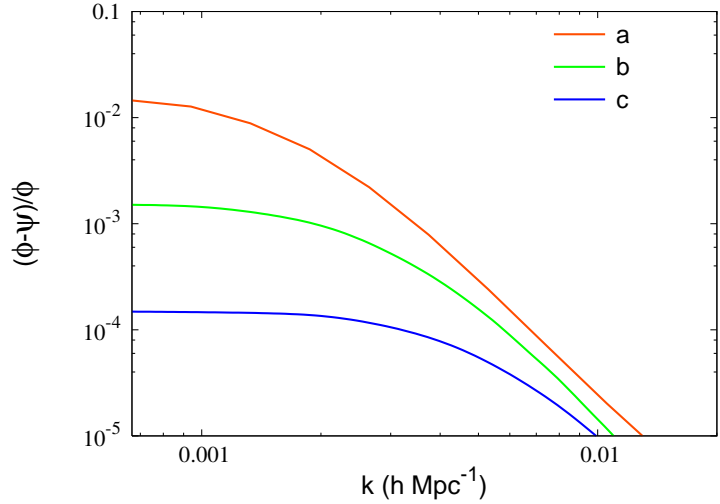}
\caption{The difference between two scalar gravitational potentials
  for several choices of parameters listed in Table~\ref{tab:1}. 
\label{Fig:6}
}
\end{center}
\end{figure}

Finally, we plot in Fig.~\ref{Fig:6} the spectrum of the relative difference
$(\phi-\psi)/\phi$ between the two gravitational potentials in
$\T$CDM. 
This 
difference is completely negligible at $k$ larger than
0.01~Mpc$^{-1}$. At smaller $k$ it can be as large as a few percent
for certain choices of parameters. 

 We stress once again that all above results are valid in the
  case of longitudinal perturbations in the Einstein-aether model
  coupled to the $\Theta$-field. $\chi$ in this case corresponds to
  the longitudinal component of the aether perturbations.

\section{Conclusions}
\label{Sec:conclusions}

In this paper we have proposed a dark energy model based on the idea
that Lorentz invariance is not an exact symmetry of nature and is
broken in the gravitational sector. As the framework for Lorentz breaking
we considered the theory of a unit time-like vector field --
aether. This can be either a general unit vector, in which case
  the setup corresponds to
 the Einstein-aether model, or it can be  
expressed in terms of the gradient
of a scalar field -- khronon -- that defines a global time coordinate. Dark
energy 
appears in this framework in a rather generic and simple way due to an
additional massless scalar with an exact shift symmetry and arbitrary
relevant couplings to the aether compatible with this invariance. An
example of such scalar is the Goldstone boson of a global broken
symmetry. 
The resulting model is a valid effective theory with a cutoff that can
be only a few orders of magnitude below the Planck mass. 
Moreover, in the khronon case the model has 
a known candidate UV completion in the form of gravity with anisotropic
scaling proposed by P. Ho\v rava.

We have shown that the  model exhibits the property of
self-acceleration. Namely, in the absence of
the cosmological constant and any matter sources the model
possesses two solutions corresponding to Minkowski and de Sitter
space-times. The former solution is unstable and the presence of an
arbitrarily small amount of matter destroys it. The cosmological
evolution of a matter-filled universe is driven to the de Sitter
attractor.
The value of the effective cosmological constant in the de Sitter
branch is determined by the lowest dimension coupling between the
Goldstone field and the aether. Importantly, it is technically natural to
assume this coupling to be small as it is protected from radiative
corrections by a discrete 
symmetry. Thus if one were able to enforce somehow the vanishing of
the vacuum energy and the existence of an (unstable) Minkowski vacuum,
the value of the current cosmic acceleration would not present
fine-tuning problems. However, we left the discussion of 
a mechanism that could lead to this cancellation of the vacuum energy
outside the scope of this article.

We analyzed the phenomenological consequences of the dark energy model
proposed in this paper. To this aim,   
we supplemented the model with cold matter and radiation
components. This particular 
realization was called $\T$CDM. We
showed that, barring fine-tuned initial conditions, the
homogeneous cosmological evolution in the model is the same as in the
universe with a cosmological constant. In other words, the effective
dark energy equation of state is $w=-1$. On the other hand, the
non-trivial dynamics of dark
energy reveals itself in the evolution of cosmological perturbations. We
studied this evolution both analytically and numerically in the linear
regime and found that the growth of perturbations is enhanced compared
to the standard $\Lambda$CDM case. The enhancement is most prominent
at very large scales of order a few gigaparsecs, but extends also to
shorter scales. Another property of the model 
is the appearance of an effective anisotropic
stress that leads at very large scales to the difference between the two 
gravitational potentials in the Newton gauge. 
In principle, the previous
effects can allow to discriminate $\T$CDM from $\Lambda$CDM. 
Several groups have recently suggested a number of data analysis
techniques to test the 
nature of dark energy and its possible deviations from the cosmological
constant  
\cite{Amendola:2007rr,Song:2010rm,Daniel:2010ky,Pogosian:2010tj,Bean:2010zq,Shapiro:2010si,Holz:2010ck,Mortonson:2010mj,Song:2010fg}.
It would be interesting to apply these techniques to the present case
and
work out the constraints on the parameters of the
model following
from the present data as well as their expected improvement by
future experiments. 

Throughout the paper we have assumed a standard Lorentz invariant 
matter sector. For the Standard Model fields this follows
from an overwhelming experimental evidence. However,  for the case of
dark matter one could relax this assumption and 
allow it to interact directly with the Lorentz breaking
fields. It would be interesting to study how these interactions may
affect the evolution of cosmological perturbations. Potentially, this
will allow to set bounds on the violation of Lorentz invariance in the
dark matter sector. We leave this investigation for the future.

\paragraph*{Acknowledgments}

We are grateful to 
Eugeny Babichev, Sergei Dubovsky, Jaume Garriga, 
Dmitry Gorbunov, Ben Hoyle, Ted Jacobson,
Julien Lesgourgues, Dmitry Levkov, Shinji Mukohyama, Jorge Nore\~na,
Oriol Pujol\`as, Valery Rubakov and Alex Vikman for illuminating
discussions. We also thank the organizers and participants of the
workshop ``Unsolved problems in astrophysics and cosmology'' for
the stimulating atmosphere during this meeting. D.B. would like to thank the 
Institute for Nuclear Research of the
Russian 
Academy of Sciences for its warm hospitality.
This work was supported in part by the Swiss Science Foundation
(D.B.), the Russian Ministry of Education and Science under the state
contract 02.740.11.0244~(S.S.),
the Grants of the President of Russian Federation
NS-5525.2010.2 and MK-3344.2011.2~(S.S.) and the RFBR grants
11-02-92108, 11-02-01528~(S.S.).

\appendix

\section{The numerical procedure}
\label{App:A}

The complete set of equations for the cosmological perturbations is 
provided by the Einstein's equations (\ref{Einslin}), equations
(\ref{khTetlin}) 
for the khronon and Goldstone perturbations 
and the hydrodynamical equations (\ref{matter}) for matter
(supplemented by the equation of state for each component).
These equations are not independent. For the numerical solution we
choose two out of the four Einstein's equations: Eqs.~(\ref{Einsij1})
and (\ref{Einsij2}). As indicated in the main text, we consider matter
consisting of two components: radiation and cold matter. Only the
former contributes into the pressure perturbation term in
(\ref{Einsij1}).
This implies that to close the system it suffices to
consider the hydrodynamic equations (\ref{matter}) only for the
radiation component. Introducing 
\[
\delta_{[\gamma]}\equiv \frac{\delta\rho_{[\gamma]}}{\bar\rho_{[\gamma]}}\;,
\]
these equations can be written as a single second order equation,
\[
\ddot\delta_{[\gamma]}-\frac{\Delta\delta_{[\gamma]}}{3}
-\frac{4\Delta\phi}{3}-4\ddot\psi=0\;.
\] 
Performing the Fourier decomposition, normalizing the
present scale factor to one, $a(t_0)=1$, and choosing the units such
that the present Hubble parameter is equal to one, $H_0=1$, we obtain
the final system of 
ordinary differential equations to be solved numerically:
\bseq
\label{eqs}
\begin{align}
\label{eq1}
&\ddot\psi+\HH(\dot\phi+2\dot\psi)+(2\dot\HH+\HH^2)\phi
-\frac{\b+\l}{2+\a B}k^2(\dot\chi+2\HH\chi)
-\frac{\Omega_\gamma}{2a^2}\delta_{[\gamma]}=0\;,\\
\label{eq1a}
&\phi-\psi-\b(\dot\chi+2\HH\chi)=0\;,\\
\label{eq2}
&\ddot\chi+2\HH\dot\chi+\big[c_\chi^2k^2+\dot\HH(1-B)+\HH^2(1+B)
+a^2c_\T^4k_c^2\big]\chi+ac_\T^2k_c^2\tilde\xi
\notag\\
&\hspace{10cm}=\dot\phi+\HH(1+B)\phi+B\dot\psi\;,\\
\label{eq3}
&\ddot{\tilde\xi}+2\HH\dot{\tilde\xi}+c_\T^2k^2\tilde\xi
+ac_\T^4k^2\chi=
-ac_\T^2\dot\phi-3a\HH c_\T^2\phi\;,\\
\label{eq4}
&\ddot\delta_{[\gamma]}+\frac{k^2}{3}\delta_{[\gamma]}+\frac{4k^2}{3}\phi
-4\ddot\psi=0\;.
\end{align}
\eseq
Here we have introduced the present radiation
density fraction $\Omega_\gamma$ and defined
$\tilde\xi=\xi/\mu^2$.   

A subtle point is the proper choice of initial conditions for the system
(\ref{eqs}).  These are fixed deep inside the radiation-domination stage
when the modes are superhorizon. We consider the initial conditions
corresponding to the adiabatic mode. The latter is regular at $t\to
0$. Thus we
write for small $t$:
\bseq
\label{expans}
\begin{gather}
\phi=\phi^{(0)}+\phi^{(1)}t\;,~~~\psi=\psi^{(0)}+\psi^{(1)}t\;,~~~
\delta_{[\gamma]}=\delta_{[\gamma]}^{(0)}+\delta_{[\gamma]}^{(1)}t\;,\\
\chi=\chi^{(0)}t+\chi^{(1)}\frac{t^2}{2}\;,~~~~
\tilde\xi=\tilde\xi^{(0)}\frac{t^2}{2}+\tilde\xi^{(1)}\frac{t^3}{6}.
\end{gather}
\eseq
Expanding Eqs.~(\ref{eqs}) at $t\to 0$ we obtain the
relations:
\bseq
\begin{gather}
\delta_{[\gamma]}^{(0)}=-2\phi^{(0)}\;,~~~~
\delta_{[\gamma]}^{(1)}=4\psi^{(1)}\;,\notag\\
\phi^{(0)}-\psi^{(0)}-3\b \chi^{(0)}=0\;,~~~~
\phi^{(1)}-\psi^{(1)}-2\b \chi^{(1)}
-\frac{\beta\Omega_{cm}}{4\sqrt{\Omega_\gamma}}\phi^{(0)}=0\;,\notag\\
\chi^{(0)}=\phi^{(0)}/2\;,~~~~
(3+B)\chi^{(1)}=(2+B)\phi^{(1)}+B\psi^{(1)}
-\frac{\Omega_{cm}}{4\sqrt{\Omega_\gamma}}\phi^{(0)}\;,\notag\\
\tilde\xi^{(0)}=-\sqrt{\Omega_\gamma}c_\T^2\phi^{(0)}\;,~~~~
\tilde\xi^{(1)}=-2\sqrt{\Omega_\gamma}c_\T^2\phi^{(1)}
-\frac{\Omega_{cm}}{2}c_\Theta^2\phi^{(0)}\;.\notag
\end{gather}
\eseq
Here we have expanded the time dependence of
the scale factor at the radiation domination epoch up to the
subleading order,
\[
a=\sqrt{\Omega_\gamma}\, t+\frac{\Omega_{cm}}{4}t^2\;.
\]
Additionally, the initial data must satisfy   
the constraint following
from the $(00)$ Einstein's equation\footnote{The $(0i)$ equation 
(\ref{Eins0i}) determines the
velocity field and does not 
impose any constraints.} (\ref{Eins00}). This gives,
\[
\psi^{(1)}+\phi^{(1)}+\frac{\delta_{[\gamma]}^{(1)}}{2}
+\frac{\Omega_{cm}}{\sqrt{\Omega_\gamma}}\left(\phi^{(0)}
+\frac{\delta^{(0)}}{2}\right)=0\;,
\]
where $\delta$ is the density contrast of cold matter,
Eq.~\eqref{eq:contrastcm}. For the 
adiabatic mode the density contrasts are related:
\[
\delta=\frac{3}{4}\delta_{[\gamma]}\;.
\]
Using this relation we obtain that all the coefficients in
(\ref{expans}) are determined in terms of $\phi^{(0)}$. 

Similar reasoning also gives the equations and initial conditions
for $\Lambda$CDM. As we are interested in the ratios
of various quantities in the $\T$CDM and $\Lambda$CDM models we can
choose arbitrary normalization for $\phi^{(0)}$, the only requirement
being that this normalization is the same for the computations in both
models\footnote{Strictly speaking, the initial spectrum in the $\T$CDM
model may be different from that in $\Lambda$CDM. Its rigorous
determination would require the analysis of the generation of
primordial perturbations at the inflationary epoch together with a model 
for the origin of the Goldstone field. These issues are beyond the
scope of the present article.
The approximation of identical
primordial spectra is enough for our purpose which is to illustrate
the difference in the evolution of the cosmological perturbations in
the $\T$CDM and $\Lambda$CDM models at the late cosmological epoch.}. 
In practice we choose $\phi^{(0)}=1$.

\end{document}